\documentclass[10pt,letterpaper]{article}
\usepackage{opex3}
\usepackage{amsmath,placeins}
\def\apjl{Astrophys. J. Lett.\ }
\def\apjs{Astrophys. J. Suppl.\ }
\def\apj{Astrophys. J.\ }
\def\aap{Astron. \& Astrophys.\ }
\def\mnras{Mont. Not. R. Astron. Soc.\ }

\begin{document}

\title{Performance characterization of a broadband vector Apodizing Phase Plate coronagraph}

\author{Gilles P.P.L. Otten,$^{1,*}$ Frans Snik,$^1$ Matthew A. Kenworthy,$^1$ \mbox{Matthew N. Miskiewicz$^{2}$} and Michael J. Escuti$^{2}$}

\address{$^1$Leiden Observatory, Leiden University, P.O. Box 9513, 2300 RA Leiden, The Netherlands\\
$^2$Department of Electrical and Computer Engineering, North Carolina State University, Raleigh, NC 27695, USA}

\email{$^*$otten@strw.leidenuniv.nl} 

\begin{abstract}
One of the main challenges for the direct imaging of planets around nearby stars is the suppression of the diffracted halo from the primary star. 
Coronagraphs are angular filters that suppress this diffracted halo. 
The Apodizing Phase Plate coronagraph modifies the pupil-plane phase with an anti-symmetric pattern to suppress diffraction over a 180 degree region from 2 to 7 $\lambda/D$ and achieves a mean raw contrast of $10^{-4}$ in this area, independent of the tip-tilt stability of the system.
Current APP coronagraphs implemented using classical phase techniques are limited in bandwidth and suppression region geometry (i.e.~only on one side of the star). 
In this paper, we introduce the vector-APP (vAPP) whose phase pattern is implemented through the vector phase imposed by the orientation of patterned liquid crystals. 
Beam-splitting according to circular polarization states produces two, complementary PSFs with dark holes on either side.
We have developed a prototype vAPP that consists of a stack of three twisting liquid crystal layers to yield a bandwidth of 500 to 900 nm. 
We characterize the properties of this device using reconstructions of the pupil-plane pattern, and of the ensuing PSF structures. 
By imaging the pupil between crossed and parallel polarizers we reconstruct the fast axis pattern, transmission, and retardance of the vAPP, and use this as input for a PSF model. 
This model includes aberrations of the laboratory set-up, and matches the measured PSF, which shows a raw contrast of $10^{-3.8}$ between 2 and 7 $\lambda/D$ in a 135 degree wedge. 
The vAPP coronagraph is relatively easy to manufacture and can be implemented together with a broadband quarter-wave plate and Wollaston prism in a pupil wheel in high-contrast imaging instruments.
The liquid crystal patterning technique permits the application of extreme phase patterns with deeper contrasts inside the dark holes, and the multilayer liquid crystal achromatization technique enables unprecedented spectral bandwidths for phase-manipulation coronagraphy.
\end{abstract}


\bibliographystyle{osajnl}

\section{Direct imaging of exoplanets}

The detection of planets around stars in our galaxy is one of the most exciting recent discoveries in astronomy. 
The first planets were not detected directly but through their effect on their parent star. 
The very first discovery of an exoplanet by \cite{Wolszczan:92} involved a planet around a pulsar, and the first exoplanet detected around a solar type star was through the radial velocity (RV) reflex motion of the parent star 51 Peg \cite{Mayor:95}. 
These massive, short-period planets implied that some of them would transit their parent star, and one such planet was found to transit its parent star, HD209458b \cite{Charbonneau:00}. 
After the discovery of this so-called Hot-Jupiter many more transiting planets have been found, with as most prolific contributor being the transit mission Kepler \cite{Borucki:09} that has detected over 2300 planetary candidates so far \cite{Batalha:13}. 
For both RV and transit searches the observing efficiency decreases sharply with increasing orbital period.
Direct imaging of extrasolar planets is required to study solar-system-like exoplanets.
In addition to direct detections, high-contrast imaging methods allow for the characterization of exoplanetary atmospheres through photometry (variability implying clouds) and spectroscopy, and are not restricted to the transit window of the planet. 
The most prominent success story for direct imaging is the 4-planet system around HR8799 \cite{Marois:08, Marois:10,Oppenheimer:13}, for which the chemical composition of HR8799 c implies a core accretion formation mechanism \cite{Konopacky:13}.

Direct imaging is highly challenging because of the large contrast between the star and the planet (typically $10^{-9}$ for visible light reflected from the Sun by Jupiter to $\sim 10^{-6}$ for direct thermal emission from young gas giant planets in formation), the small angular separation between the star and the planet (typically 1 arcsecond or less), and the effects of diffraction of the stellar light due to the finite size of the telescope. 
The diffraction halo of the primary star (the point spread function - PSF) at the location of the planet can be several decades brighter than the planet itself. 
Contrast from the ground is furthermore limited by a halo of speckles created by imperfect Adaptive Optics (AO) correction of atmospheric turbulence and optical aberrations that are not corrected by the AO system as they are not sensed by the wavefront sensor.
Removing the diffraction halo by telescope modeling or observing a nearby reference star is not effective because the optics of the telescope changes as a function of time, and hence the resulting PSF is variable at the pertinent levels of contrast. 
The speckles that result from incorrect PSF subtraction are similar in shape to the planets that are the target of our search. 
Several differential techniques for improving PSF subtraction have been developed and are used to obtain roughly an additional order of magnitude in suppression: Angular Differential Imaging \cite{Marois:06, Lafreniere:07, Amara:12, Meshkat:14}, Spectral Differential Imaging \cite{IRDIS}, Spectral Deconvolution \cite{SparksFord} and Polarimetric Differential Imaging for scattered light sources (see \cite{Avenhaus:14} for a recent example).

The intensity of the speckles is related to the telescope wavefront error and the intensity of the diffraction structure at the location of the planet \cite{Bloemhof:03}. 
By suppressing the diffraction structure from the central star the effect of the intensity of the corresponding speckles is also reduced. 
Coronagraphs are angular filters designed to reject light from the direction of the star while transmitting as much light as possible from the direction of the planet. 
Originally developed for the study of the solar corona \cite{Lyot:39}, several different designs have been considered for use in direct imaging of exoplanets (for recent comprehensive reviews see \cite{Guyon:05} and \cite{Mawet:12}). 
New-generation coronagraphs such as the Apodizing Phase Plate and Annular Groove Phase Mask coronagraph have been in use at near-infrared wavelengths in ground-based telescopes \cite{Quanz:10,Mawet:13}. 
Out of the wide family of coronagraphic designs, those which use focal-plane amplitude and/or phase masks to suppress the on-axis starlight are sensitive to both the finite angular size of the star and tip-tilt vibrations within the telescope structure. 
Coronagraphs which modify the complex amplitude of the pupil plane only are not susceptible to these problems. 
Modifications of the pupil through amplitude apodization \cite{Kasdin:04,Carlotti:13} and phase apodization \cite{Codona:04,Yang:03,Yang:04,Guyon:05b} are possible and can provide a combination of large suppression of starlight and high throughput of the light of a planet at small angular separations from the central star. 
In \cite{Yang:04} the theory of pupil apodization in one dimension was developed and extended for square apertures. 
Pupil apodization over circular apertures was further developed in \cite{Kostinski:05} with iterative Gerchberg-Saxton algorithms, clearing out regions in the focal plane at typically 5 $\lambda/D$ and larger. 
\cite{Codona:04} and \cite{Kenworthy:07} developed and realized phase coronagraphs at smaller inner working angles. 
The largest advantages of pupil-plane over focal-plane coronagraphs are the small inner working angle, easy accommodation of complex pupil shapes (segments/spiders) and its insensitivity to tip-tilt variations. 
One disadvantage of a stand-alone pupil-plane coronagraph is that all the stellar light still reaches the detector, which leads to dynamic range issues.

The original Apodizing Phase Plate (APP) \cite{Kenworthy:07,Codona:06} introduces a phase variation across the telescope pupil by varying the thickness of a plate as a function of position in the pupil, and has been used successfully to directly image exoplanets (e.g. \cite{Quanz:10}). 
For the APP these thickness variations are typically diamond-turned into a substrate with a high refractive index. 
The two limiting factors for this APP design are (1) it is inherently chromatic and (2) it only delivers 180 degrees field of suppression. 
However, the implementation of phase control through the \emph{vector phase} solves these two problems, and enables an improved version of the APP coronagraph, which is the subject of this paper. 
The much improved performance for the vector Apodizing Phase Plate (vAPP) is achieved with a combination of liquid crystal techniques: 
The phase pattern can be written in a photo-alignment material \cite{Yaroshchuk:12} with a direct-write scanning system \cite{Miskiewicz:14}.
Broadband performance is achieved by applying multiple layers of self-aligning liquid crystals (multi-twist retarders; MTR), that together yield an achromatized retardance \cite{Komanduri:13}, which extends the efficiency of the inherently achromatic vector-phase application over a large wavelength range.

In this paper we present the characterization of the first vAPP prototype, which creates dark holes in two complementary PSFs over a broad wavelength range (500--900 nm). 
In Section \ref{sec:2} we describe the principle behind the vAPP.
In Section \ref{sec:3} we directly characterize the manufactured optic through narrowband imaging at multiple wavelengths. 
Using these measurements we then model the PSF in the focal plane in Section \ref{sec:4} and compare it with measurements of the PSF in the laboratory.
We present our conclusions and provide outlooks for future improvements in Section \ref{sec:5}.

\section{vAPP operating principle \& manufacturing}
\label{sec:2}

\subsection{Classical \& vector phase}
\label{sec:geovecphase}

\begin{figure}[!ht]
\centerline{\includegraphics[scale=0.35]{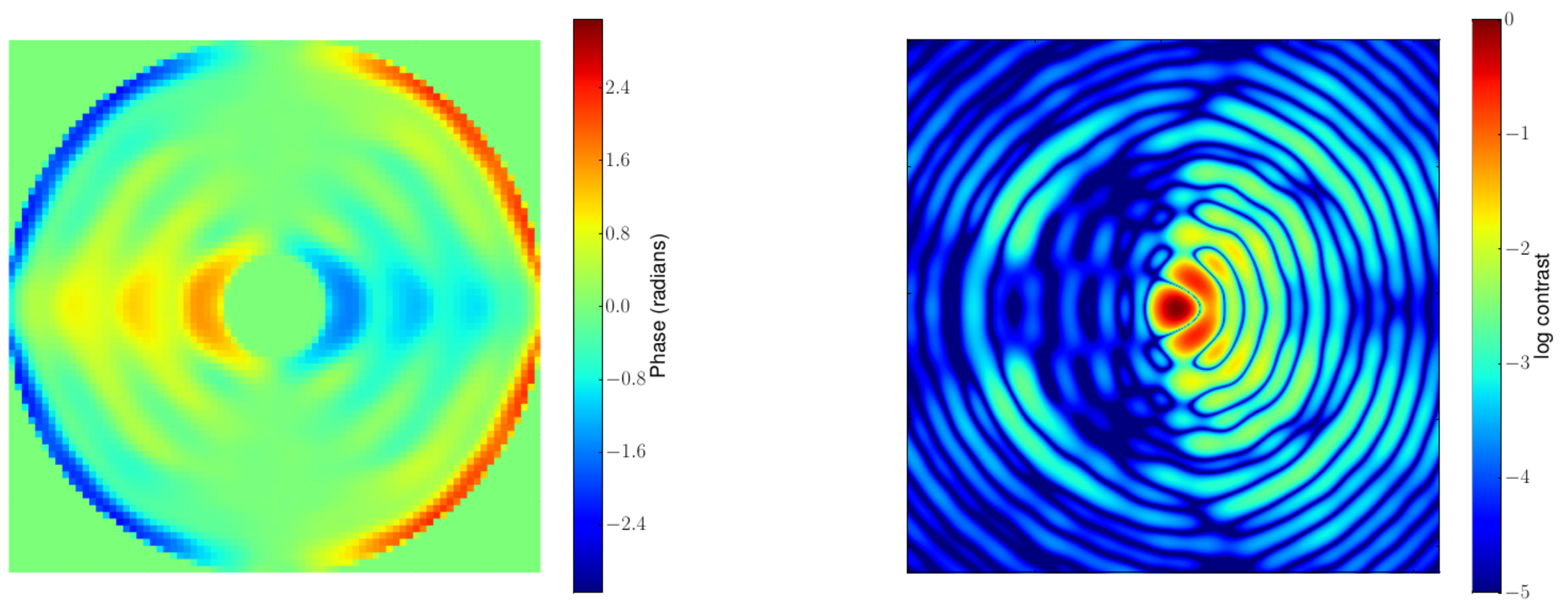}}
\caption{Left panel: The anti-symmetric phase pattern of the Apodizing Phase Plate coronagraph that is used in the vAPP prototype. Right panel: Corresponding theoretical log-scaled PSF, normalized to its peak flux.}
\label{fig:app_phase}
\end{figure}

The APP coronagraph uses optical path differences to induce a phase change as a function of position across the pupil. 
One such phase pattern is shown in Fig. \ref{fig:app_phase}.
In this case, a phase pattern $\Delta\phi$ is introduced by varying the thickness $\Delta x$ of a high refractive index material. 
The expression relating $\Delta\phi$ to $\Delta x$ is given in Eq. \eqref{eq:opd} where $\lambda$ is the wavelength, $n_\mathrm{plate}$ is the refractive index of the coronagraph and $n_\mathrm{air}\approx 1$.

\begin{equation}
\Delta\phi = \frac{2\pi}{\lambda} (n_\mathrm{plate}(\lambda)-n_\mathrm{air}(\lambda)) \Delta x
\label{eq:opd} 
\end{equation}

This phase is clearly chromatic with a $1/\lambda$ term and a dispersion term for the glass.

The vAPP coronagraph instead makes use of the geometric phase, also sometimes called the vector phase, which is a manifestation of the Pancharatnam-Berry phase \cite{Pancharatnam:56,Berry:87}. 
This approach was previously used for focal-plane phase masks like the 4QPM (with quartz/$\mathrm{MgF}_2$ achromatic waveplates) by \cite{Mawet:06}, the 8OPM (with photonic crystals) by \cite{Murakami:10}, the Vector Vortex Coronagraph by \cite{Mawet:09}, and introduced for the vAPP by \cite{Snik:12}. 
The vector phase arises when a circularly polarized beam passes through a half-wave retarder and is converted into circularly polarized light of the opposite handedness.
While the circular polarization handedness flip is independent from the half-wave retarder's axis orientation, the absolute phase of the emergent beam is directly determined by it. 
In this way, even a perfectly flat optic can induce a phase pattern. 
Moreover, as the phase is only determined by the fast axis orientation, it is inherently achromatic.

The electric components of a beam of polarized light can be described by a Jones vector \mbox{$\mathbf{E}=\left(\begin{matrix}  E_x  \\  E_y\end{matrix} \right)$}.
Using the Jones formalism to express the effect of a half-wave plate (HWP) with a fast axis oriented at $\theta_\mathrm{HWP}$ on circular polarization states we can write 
\begin{equation}
\mathbf{E_\mathrm{out}}=\mathbf{J_\mathrm{rot}}(-\theta_\mathrm{HWP})\,\mathbf{J_\mathrm{HWP}}\,\mathbf{J_\mathrm{rot}}(\theta_\mathrm{HWP})\,\mathbf{E_\mathrm{in,circular,\pm}}
\label{eq:jones}
\end{equation}
where the rotation matrix is
\begin{equation}
\mathbf{J_\mathrm{rot}}\left(\theta\right)=\left( \begin{matrix} 
\cos \theta & \sin \theta \\ 
 -\sin \theta & \cos \theta
\end{matrix} \right),
\end{equation}
the Jones matrix of the HWP
\begin{equation}
\mathbf{J_\mathrm{HWP}}=\left( \begin{matrix} 
\exp\left(-i\, \pi/2\right) & 0 \\ 
 0 & \exp\left(i\, \pi/2\right)
\end{matrix} \right)\,.
\end{equation}
Adopting circularly polarized input states \begin{equation} \mathbf{E_\mathrm{in,circular,\pm}}=\frac{1}{\sqrt{2}}\left(\begin{matrix}  1  \\  \pm i\end{matrix} \right)\label{eq:circpol}\end{equation} gives us 
\begin{equation}\mathbf{E_\mathrm{out}}=-\frac{i}{\sqrt{2}}\left( \begin{matrix} 1  \\ \mp i \end{matrix} \right) \, \exp(\mp i 2\,\theta_\mathrm{HWP}) \,, \end{equation} 
which shows that the input circular polarization has flipped its sign and received a phase delay equal to
\begin{equation}
\Delta\phi = \mp 2\,\theta_\mathrm{HWP}.
\label{eq:vector} 
\end{equation}
It can be seen that the expression of the vector phase has a positive or negative phase depending on the handedness of the circularly polarized light, and is independent of wavelength.
If the retarder is not perfectly half-wave, the handedness-flip is incomplete and leakage terms emerge \cite{Mawet:09}.

\begin{figure}[!ht]
\centerline{\includegraphics[scale=0.85]{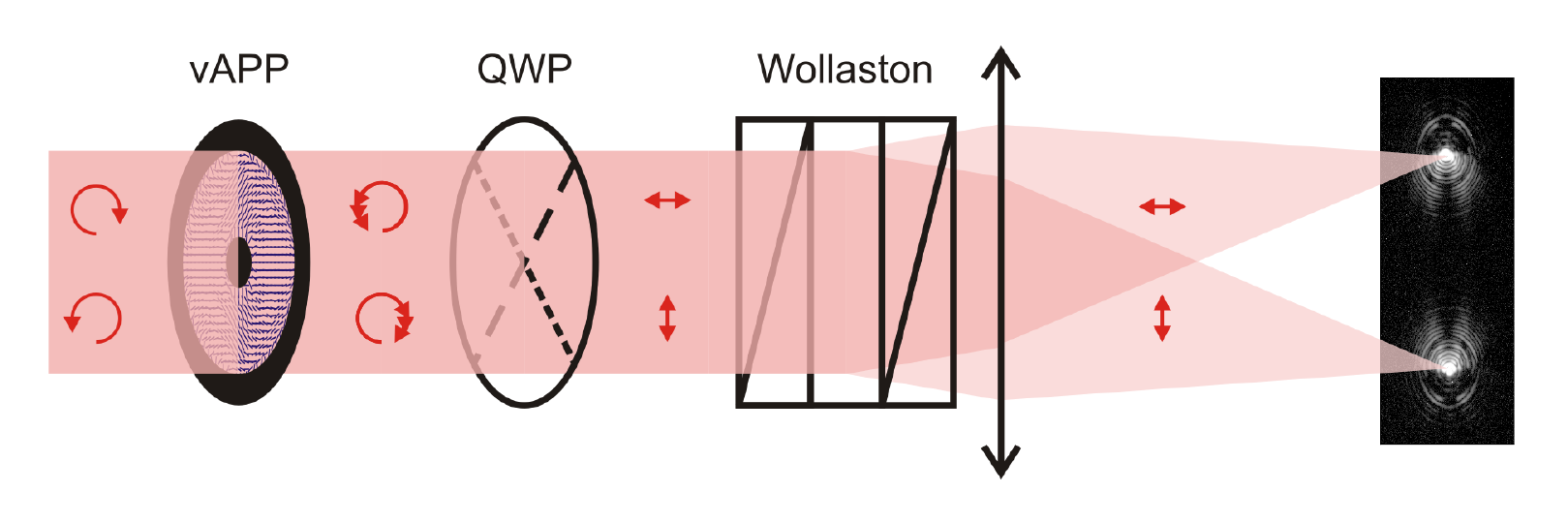}}
\caption{The operational principle of the vector Apodizing Phase Plate coronagraph. The vAPP is a patterned half-wave plate that converts input circular polarizations into their opposite handedness with an extra phase delay added. One circular polarization accrues a positive phase pattern while the opposite polarization receives a negative phase. A QWP plate at 45 degrees converts these circular into linear polarization states. The Wollaston prism then splits the beam based on the linear polarizations, while a camera lens images complementary PSFs on the camera. The log-scaled grey-scale image on the right-hand side of the image shows an example of these PSFs as measured in the lab.}
\label{fig:vapp_principle}
\end{figure}

Unpolarized light can be decomposed into equal components of the orthogonal circular polarization states, and therefore half of the light accrues a positive phase while the other half receives a negative phase. 
Since the APP phase pattern is anti-symmetric, the negative phase corresponding to opposite circular polarization yields a point-reflected PSF compared to the PSF of the positive phase. 
This can be easily understood by looking at the properties of the Fourier transform. 
The sign flip of the phase pattern is equivalent to the complex conjugation of the electric field pattern in the pupil.
The Fourier transform of the two pupil functions yields electric field structures in the focal plane that are point-reflected ($[x,y]\rightarrow[-x,-y]$) complex conjugates.
The PSFs (square of the modulus of the electric field pattern in the focal plane) for the two circular polarization states therefore exhibit point-reflection symmetry.
In the case of an anti-symmetric pupil phase pattern, the two PSFs are mirror-symmetric.
Therefore, if the light is split based on the handedness of the emerging beam we obtain two distinct images of the same source but with opposite sides cleared out. 
Figure \ref{fig:vapp_principle} shows a sketch of the vector APP coronagraph. 
In this case, the beam-splitting according to circular polarization states is achieved by introducing a quarter-wave plate (QWP) and Wollaston prism behind the vAPP.
The QWP converts the circular polarizations into linear polarizations that are split by the Wollaston prism.

We use liquid crystal techniques to create the desired pattern retarder, which has a retardance close to half-wave over a large wavelength range.
While a single layer of in-plane, uniaxial nematic liquid crystal formed into the vAPP pattern produces output waves with phase following Eq. \eqref{eq:vector}, it nevertheless has a very chromatic retardance, i.e., proportional to $1/\lambda$. 
This means that only a narrow band of wavelengths will experience a half-wave retardance, even approximately. 
At other wavelengths, the large deviations from half-wave retardance create leakage of the original PSF. 
An achromatized half-wave retarder can be created by coating multiple layers of chiral liquid crystals, each with its own thickness and helical twist \cite{Komanduri:13}. 
This static thin film with multiple sublayers is called a Multi-Twist Retarder (MTR). 
In addition to its broadband performance, the other key advantage of this birefringent film is that the local fast axis of each layer spontaneously self-aligns to the fast axis of the prior layer, implying that only one patterning step for a single photoalignment material is needed. 
Furthermore, in this work, we direct-write the vAPP fast axis orientation pattern using a scanned laser spot, a direct-write lithography technique \cite{Miskiewicz:14}, which has the advantageous feature of being able to record \emph{any} phase pattern. 

\subsection{Fabrication of the vAPP prototype}

The vAPP prototype characterized in this paper is made on a substrate (Schott D263 glass) with a diameter of 25.4 mm and a thickness of 1.1 mm. 
The first layer applied to the glass is a linearly photo-polymerized polymer (LPP \cite{Yaroshchuk:12}), which defines the local optical axis orientation for subsequent layers and therefore the fast axis of the whole device. 
A linearly polarized UV laser (325 nm) encodes the APP phase as a position-dependent fast axis orientation pattern into the LPP layer with a spatial resolution of 25 microns. 
The angle of polarization varies with a rotating half-wave plate \cite{Miskiewicz:14} in order to locally enforce the  polymerization direction to follow the required phase pattern (see Fig. \ref{fig:app_phase}). 
Note that a factor of two in angle is applied to the phase pattern, cf.~Eq.~\ref{eq:vector}.
After the curing of the LPP, successive layers of liquid crystal form the achromatized, patterned half-wave retarder.
The multilayer design compensates for the chromaticity of individual layers by a combination of varying thicknesses, twists and birefringences. 
For our prototype, we use three total layers (i.e., a 3TR design \cite{Komanduri:13}) of liquid crystal polymers and cure each layer with a UV LED source. 
The properties of the individual MTR layers are given by [$\theta_1=70^\circ$, $d_1=1.45\,\mu$m, $\theta_2=0^\circ$, $d_2=3\,\mu$m, $\theta_3=-70^\circ$, $d_3=1.45\,\mu$m] where $\theta_i$ denotes the local fast axis orientation and $d_i$ the thickness of each layer. 
In more detail; the process involves spin-coating successive layers of a mixture containing polymerizeable liquid crystal monomers and potentially chiral dopants within a solvent, and results in a solid cross-linked polymer film.
Further details of the design method and fabrication principles are found in \cite{Komanduri:13}.
An end-cap glass plate with a broadband anti-reflection coating is added. 

The multilayer liquid crystal structure creates a retardance that is designed to be approximately half-wave between 500 and 900 nanometers. 
To minimize the leakage of the regular PSF to a level comparable to the theoretical contrast of the coronagraph we specify a maximum retardance offset. 
By comparing the contrast in a 135 degree wedge from 2 to 7 $\lambda/D$ for the normal PSF and the theoretical APP PSF for the adopted phase pattern we see that we can tolerate a leakage of 7.3\%. 
The intensity of the leakage term as a function of the retardance offset $\Delta\delta$ is approximately $\sin^2\left(\Delta\delta/2\right)$ \cite{Snik:12}, and from this we derive that the maximum retardance offset that can be tolerated is $\Delta\delta=0.55\ \mathrm{radians}$. 
The MTR design for the vAPP prototype is selected to fall within this requirement (see Fig.~\ref{fig:measured_retardance}).
A coronagraph design with improved contrast will have a stronger requirement on the retardance offset. 
For instance, a design with a 10 times better theoretical contrast requires a maximum retardance offset of 0.17 radians, and/or the application of leakage filtering with circular polarization filters on either side of the vAPP \cite{Mawet:09, Snik:14spie}.

\begin{figure}[h]
\centerline{\includegraphics[scale=0.25]{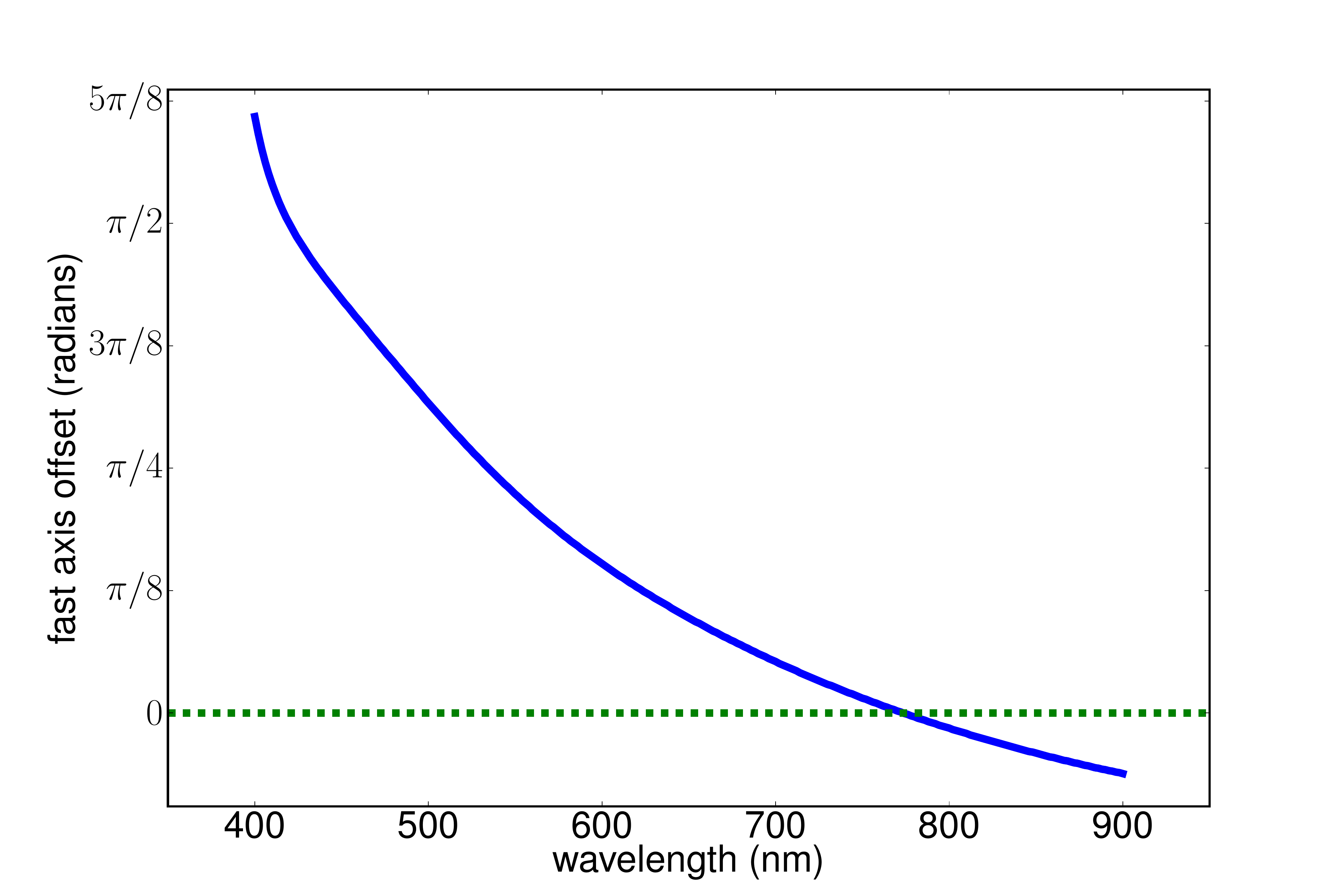}}
\caption{Theoretical fast axis offset as a function of wavelength for the vAPP prototype. This apparent fast axis rotation is derived from the theoretical MTR Mueller matrix, and is used to offset the polarizers in the measurements of the phase pattern.}
\label{fig:fastaxisoffset}
\end{figure}

\begin{figure}[h]
\centerline{\includegraphics[scale=0.3]{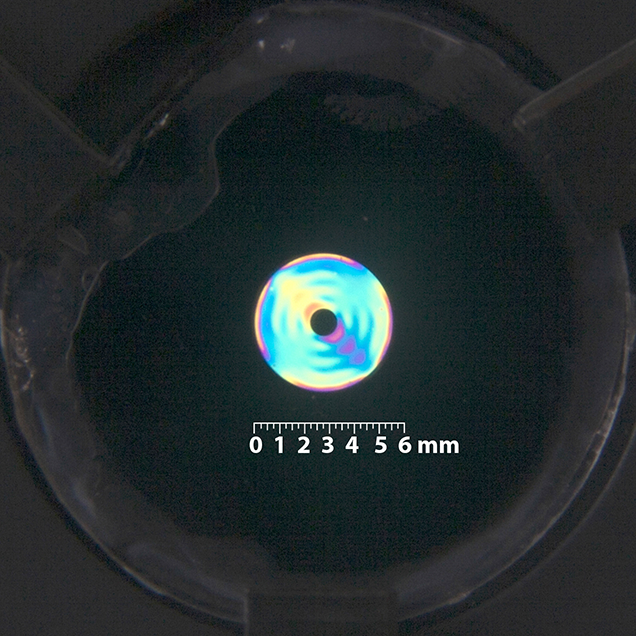}}
\caption{The vAPP prototype seen between crossed polarizers. The colors are due to the global rotational offset of the fast axis as a function of wavelength.}
\label{fig:vapp_plate}
\end{figure}

While the MTR design used for the vAPP prototype is optimized to create an approximately half-wave retardance across a broad wavelength range, the fast axis zero point is left unconstrained. 
The effect of the zero point is a global rotation of the fast axis and therefore a harmless global piston offset of the phase of the emergent beam as a function of wavelength. 
The dependance of the zero point on the wavelength is derived from the theoretical 3-layer MTR Mueller matrices and is shown in Fig. \ref{fig:fastaxisoffset}.
This feature creates a paradoxically colorful appearance when viewing the broadband vAPP prototype in between polarizers (see Fig.~\ref{fig:vapp_plate}).

An amplitude mask is made using a substrate with the same specification as is used for the patterned liquid crystal. 
A layer of opaque (black) resist, with a transmissive annulus with an inner diameter of 1.1 mm and outer diameter of 5.5 mm is applied to this substrate. 
The transmission of this mask is measured to be lower than $10^{-5.5}$. 
This amplitude mask is manually aligned to the other substrate under a microscope and bonded with optical glue. 
The completed optic can be seen in Fig. \ref{fig:vapp_plate}.

\section{Measuring the properties of the vAPP}
\label{sec:recon}
\label{sec:3}
Detailed characterization of the broadband optical behavior (i.e., transmission, retardance, and fast axis) of the vAPP is necessary not only to verify if the plate is functioning according to specifications, but also to identify solutions to perfect the current design and reach the performance needed to accommodate more complex coronagraphs.
The vAPP prototype is inserted in between rotatable polarizers, and a lens re-images the pupil plane onto a detector, see Fig.~\ref{fig:pup_setup}. For details about this setup see Appendix~\ref{sec:appa2}.
By recording images for different (crossed, parallel) configurations of the polarizers, and application of a Mueller matrix model (see Appendix \ref{sec:appa1}), the transmission, retardance and fast axis patterns are derived for a range of narrow-band filters from 500 to 800 nm.

\begin{figure}
\centerline{\includegraphics[scale=0.85]{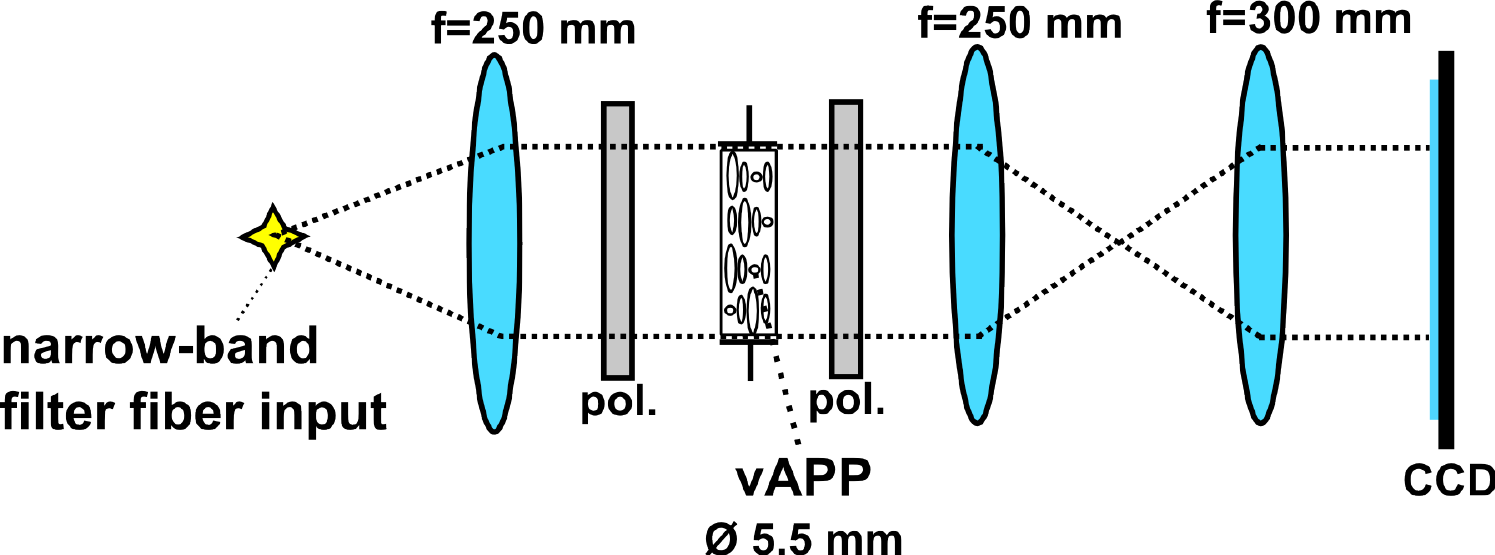}}
\caption{Layout of the pupil measurement setup of the vAPP. The vAPP is placed between linear polarizers and its surface is re-imaged onto a CCD. The intensity images of the vAPP taken at different polarizer angles are used to reconstruct the fast axis, retardance and transmission of the optic as described in Appendix \ref{sec:appa1}. The setup is described in more detail in Appendix \ref{sec:appa2}.}
\label{fig:pup_setup}
\end{figure}

\begin{figure}[h]
\centerline{\includegraphics[scale=1.05]{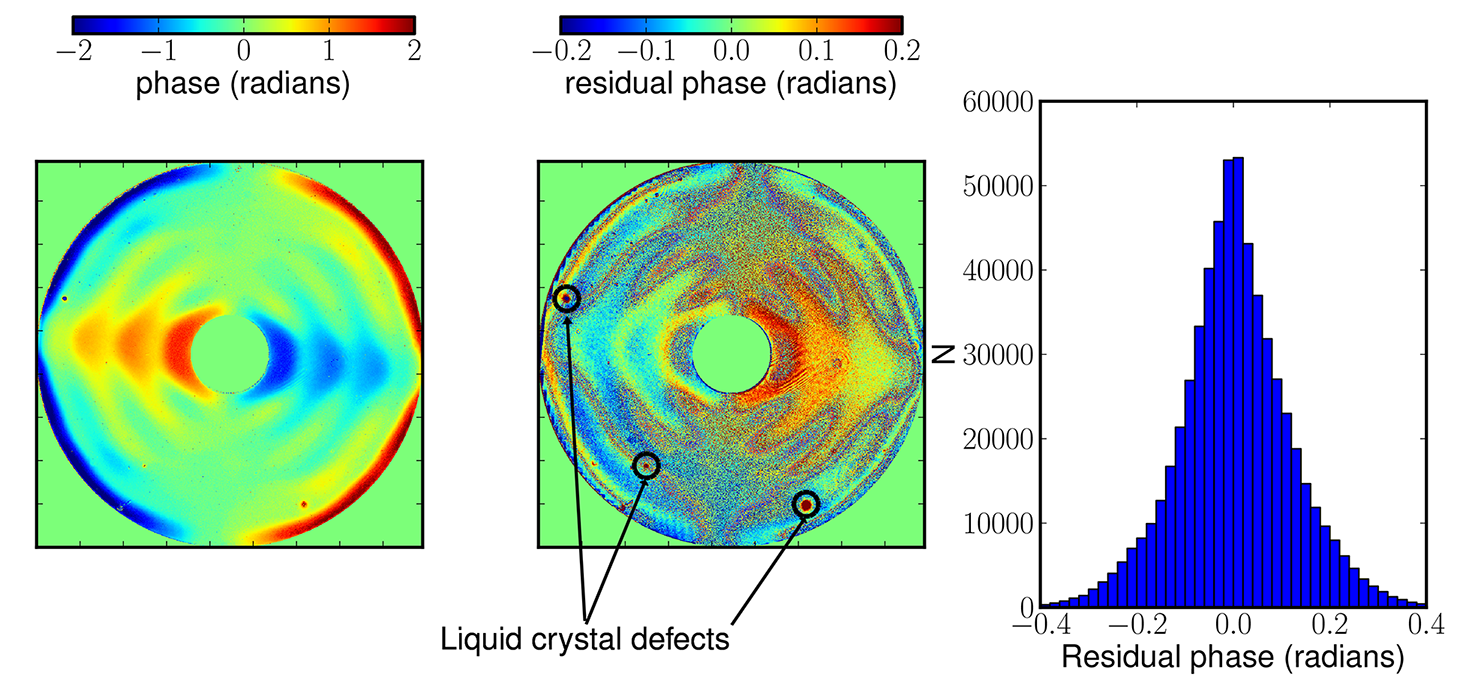}}
\caption{Left panel: Reconstructed phase (i.e.,~two times the retarder axis orientation) map of vAPP. Middle panel: Difference between reconstructed phase of vAPP and target phase pattern used for manufacturing. Small irregularities can be seen in the map. Some regions of the plate have a peculiar phase pattern due to dust specks present on the substrate during manufacturing. The dust specks locally create a thicker layer of liquid crystals. Right panel: Histogram of phase residuals.}
\label{fig:phase_map}
\end{figure}

The measured phase pattern at 750 nm, and the difference between the reconstructed phase map of the pupil and the input phase map of Fig.~\ref{fig:app_phase} can be seen in Fig.~\ref{fig:phase_map}. 
The reconstructed phase map (i.e.~two times the retarder axis orientation pattern) is remarkably similar to the input pattern although small, local deviations can be discerned.
Especially at the 270 degree and 150 degree clock-wise positions two peculiar phase deviations can be seen. These are due to dust specks that landed on the substrate during manufacturing and locally created a thicker layer of polymers. 
The pile-up of polymers causes an offset in retardance and phase.

The non-zero residuals in the phase shown in Fig. \ref{fig:phase_map} are partially caused by measurement errors. 
The fuzzy edges that trace out some of the phase pattern are evidence of this. 
The phase folds back at the extremes of the arccosine function in Eq. \eqref{eq:fastaxis}, where the derivative of the arccosine is large and a small change in the ratio can cause a big difference in phase. 
Measurement noise is therefore very pronounced at these locations. 
Furthermore, there is a small phase jump at this location. 
Simulations of the end-to-end measurement trace this phase jump back to a 1\% residual background value in the original images. 
The use of a (well-calibrated) complete Mueller matrix imaging polarimeter would reduce such issues, although some degeneracies would remain.
In addition to these spurious effects, the phase pattern appears to have a small ($\ll$0.1 radians) real deviation from the specified pattern (see Fig. \ref{fig:zigzag} in the Appendix).
This may be caused by instrumental polarization issues in the direct-write system.

\begin{figure}[h]
\centerline{\includegraphics[scale=0.31]{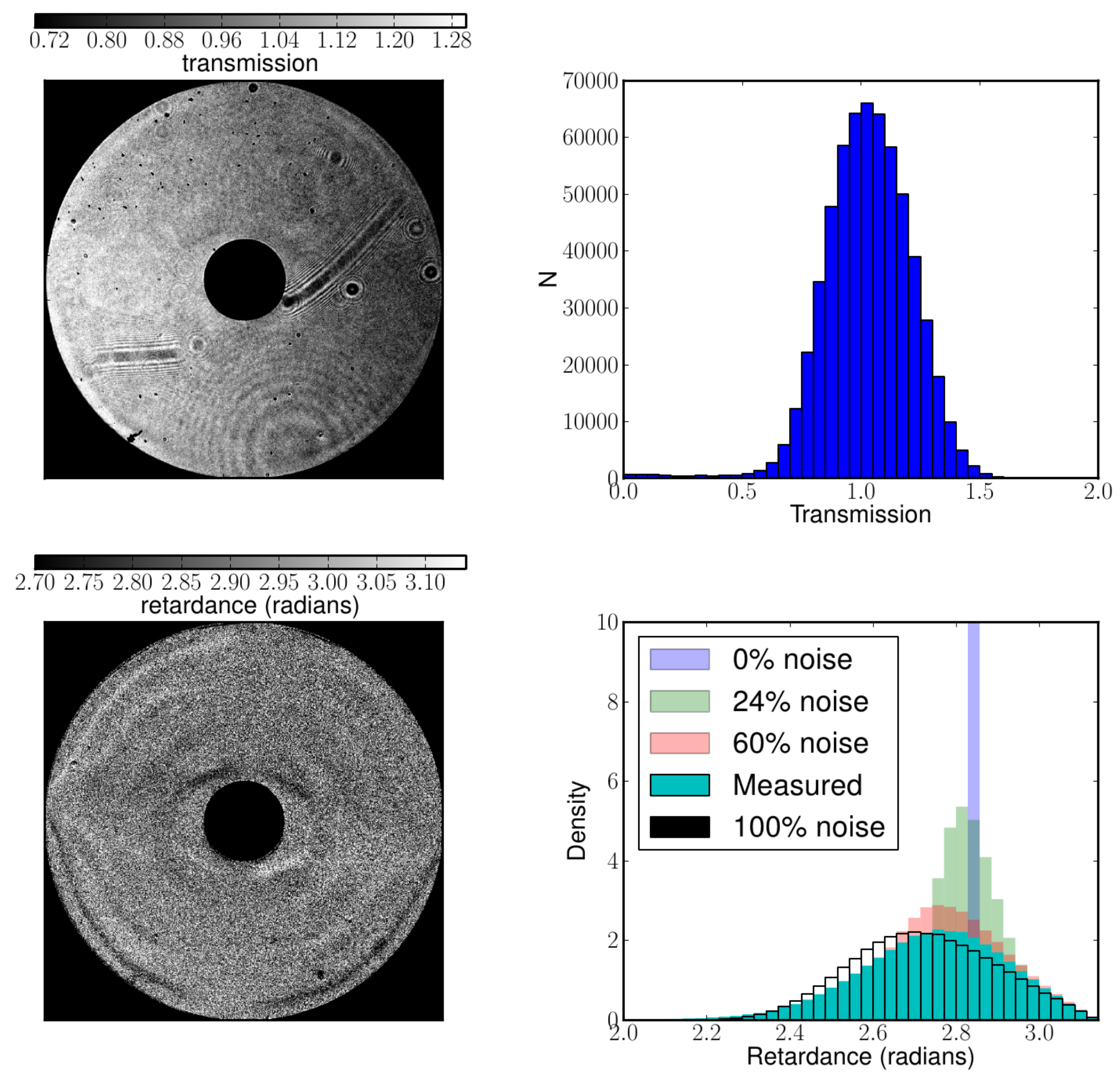}}
\caption{Upper panel: Measured transmission map at 750 nm. The transmission includes effects caused by the optical setup. Lower panel: Reconstructed retardance map at 750 nm. The right hand side shows the corresponding histograms. The histogram of the retardance also shows the expected simulated histograms assuming a retardance that matches the theoretical value given by the MTR design. Different amounts of Gaussian noise were injected, where 100\% corresponds with the measured noise in the recorded images. This noise offsets the peak and skews the distribution.}
\label{fig:ret_trans_map}
\end{figure}

The retardance and transmission maps are shown in Fig.~\ref{fig:ret_trans_map} together with the corresponding histograms. 
A hint of the phase pattern can be seen in the retardance map, because it is reconstructed simultaneously with the phase map.
Simulations show that this is caused by a shift of a few pixels between the four recorded images that are used to compute this map.  
This shift is caused by the different orientations of the polarizers that are required to derive the retardance and phase maps (see Appendix \ref{sec:appa1}).
The polarizer is slightly tilted on purpose to remove reflection ghosts that would otherwise bias the characterization.
Furthermore, the average of the measured retardance map is offset from the theoretically expected value for the 3TR design.
Simulations using the Mueller matrix model in Appendix \ref{sec:appa1} show that this is caused by the propagation of noise in the recorded images.

The transmission map shows a gradient across the pupil but does not show any print-through of the phase pattern.
A global gradient of light going from a flux ratio of 1.25 to 0.75 across the pupil is seen in the topleft panel of Fig. \ref{fig:ret_trans_map}. 
To investigate the cause of this gradient we divide the APP pupil image by a flat field image created by removing the vAPP optic. 
Inspection of this image shows no gradients in transmission across the APP optic to within 1\%. 
We conclude that the gradient seen in the pupil transmission image is caused by an asymmetric illumination of the pupil.

Amplitude variations can be seen at several locations on the pupil image. 
The long black fringes are scratches in the cover window of the CCD and the circular fringes are from dust specks on the lenses located close to the vAPP. 
These effects can be corrected by improving the optical setup. 
Other amplitude effects are in the coronagraph itself and have to be solved in the manufacturing step.
For instance, the small circles on this image are air bubbles and dust specks in the glue that binds the amplitude mask to the coronagraph.

Finally in the lower left of the coronagraph image a small indentation is seen where the transmission is substantially lower. 
This seems to be a patch of resist that is sticking out of the amplitude mask. In the phase and retardance reconstruction the static amplitude variations are removed automatically. 
In the PSF reconstruction presented in Sect.~\ref{sec:4} it is shown that these amplitude variations do not have a significant impact on the performance of the coronagraph.

\begin{figure}[!h]
\centerline{\includegraphics[scale=0.3]{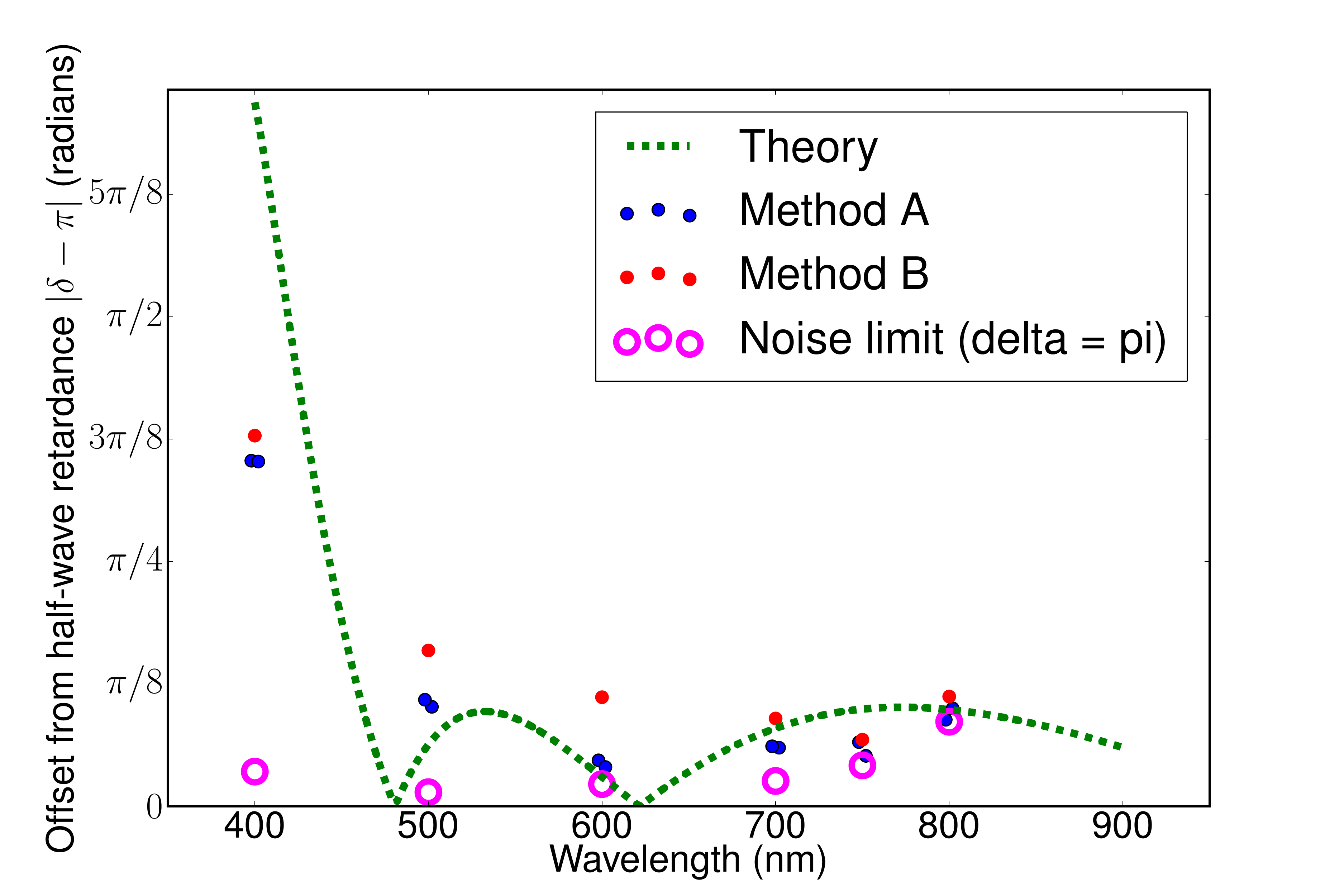}}
\caption{Measured offset of the retardance from $\pi$ radians as a function of wavelength. Denoted in red and blue circles are the two different reconstruction methods. Method A uses 4 pairs of images to solve for both the fast axis and the retardance while method B only looks the antisymmetry axis in pairs of images. Overplotted in a dashed green line is the theoretical retardance curve for a 3-layer MTR. The pink circles represent the noise limit of the retardance measurement using the noise and flux of the real images to retrieve the retardance of a HWP that is exactly half-wave. At 750 and especially 800 nm it is seen that these dots roughly overlap with these limits which shows these values are only an upper limit.}
\label{fig:measured_retardance}
\end{figure}

Figure~\ref{fig:measured_retardance} presents the measured offsets from half-wave retardance as a function of wavelength for the two different reconstruction techniques detailed in Appendix \ref{sec:appa1}.
The retardance values are measured as an average across the entire device. 
Because the retardance reconstruction in Eq. \eqref{eq:retardance} has a degeneracy, we plot the offset from half-wave retardance, which is the factor that determines leakage of the original PSF.
The 3-layer design approximates half-wave retardance for the specified spectral range of 500--900 nm to within 0.3 radians or 17.2 degrees.
The measurements of retardance as a function of wavelength well match the expected performance.
The purple dots in Fig. \ref{fig:measured_retardance} represent the systematic error offset due to the photon noise (see Appendix \ref{sec:appa3}), and are a lower bound for the measurement error of the retardance offset. 
The large systematic error at 800 nm is caused by the very low photon efficiency of the used detector.
At 750 nanometers both method A and B give significantly lower retardance offsets than the prediction.
At 500 and 600 nm, the two employed methods do not agree, which is likely caused by fluctuations in the light source.

\begin{figure}[!h]
\centerline{\includegraphics[scale=0.85]{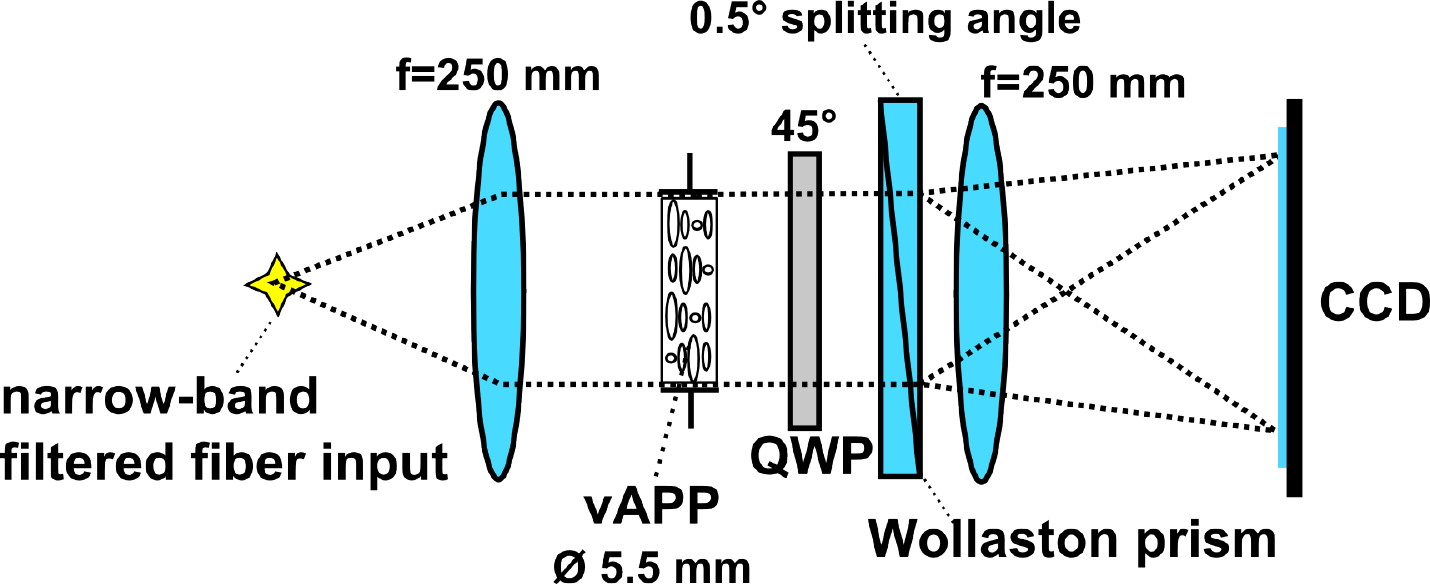}}
\caption{Layout of the PSF measurement setup. The working principle is the same as seen in Fig. \ref{fig:vapp_principle}. The vAPP is re-imaged onto the detector by a lens of $f=250\,\mathrm{mm}$. By splitting the two  circular polarizations with a QWP at 45 degrees and a Wollaston prism with a 0.5 degree splitting angle, two PSF spots with complementary dark holes are recorded in the focal plane.}
\label{fig:schematic_psf}
\end{figure}

\section{PSF characterization}
\label{sec:4}
In directly imaging exoplanets, where the halo of the primary star dominates the flux at the location of the planet, it is important to know the intensity as a function of angular separation away from the central star. 
This is expressed in an azimuthally averaged radial light profile called a contrast curve. 
The contrast curve is typically normalized to unity at the location of the star and the angular separation expressed in units of diffraction width $\lambda/D$, with $D$ the diameter of the pupil aperture. 
From this contrast curve the sensitivity of a coronagraph for a given observation can be calculated. 
We therefore model the theoretical PSF of the vAPP as a function of wavelength based on the results of the pupil-plane measurements presented in the previous Section and compare it to laboratory PSF measurements.

\begin{figure}[!htb]
\centerline{\includegraphics[angle=0,origin=c,scale=0.575]{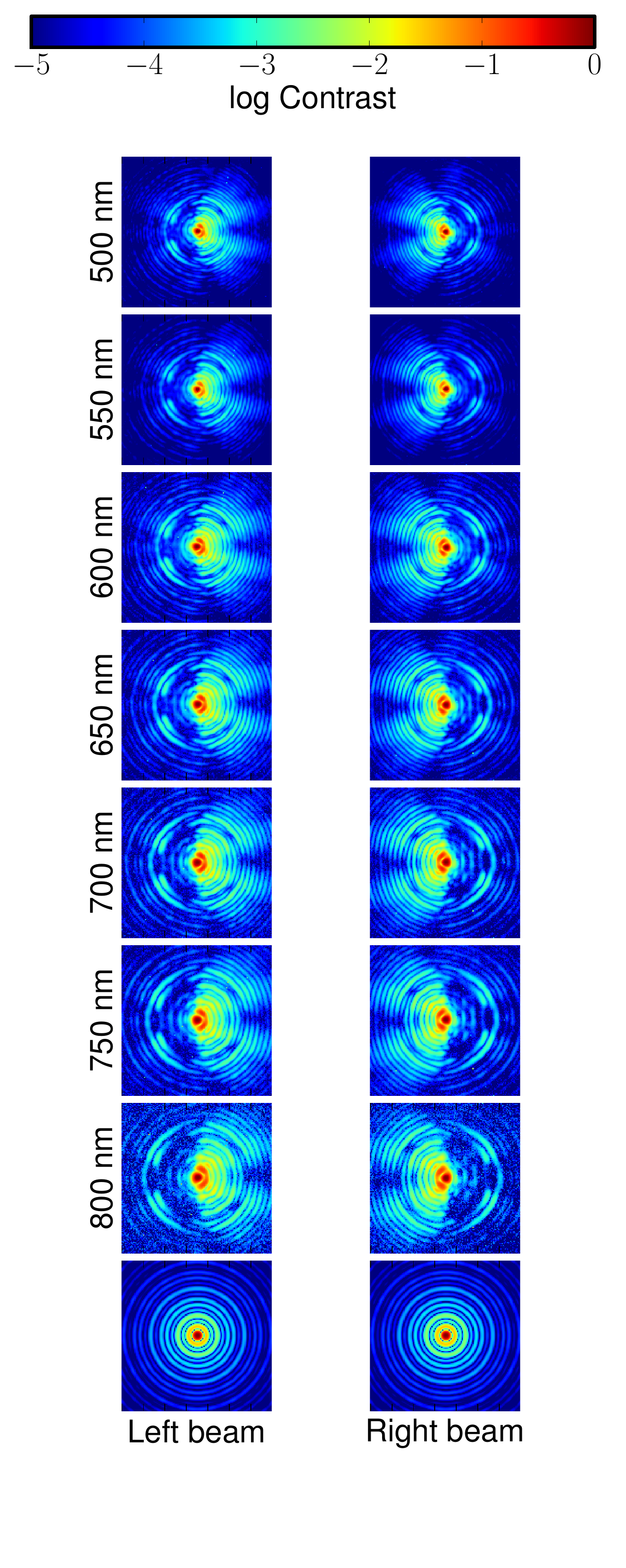}}
\caption{Laboratory measurements of PSFs ranging from 500 to 800 nm showing the PSFs of the Left and Right beam. For comparison, the bottom two panels  show the unaberrated PSFs at 800 nm. All PSFs are normalized to the peak of the PSF and scaled logarithmically.}
\label{fig:allpsfs}
\end{figure}

\begin{figure}[!ht]
\centerline{\includegraphics[scale=.35]{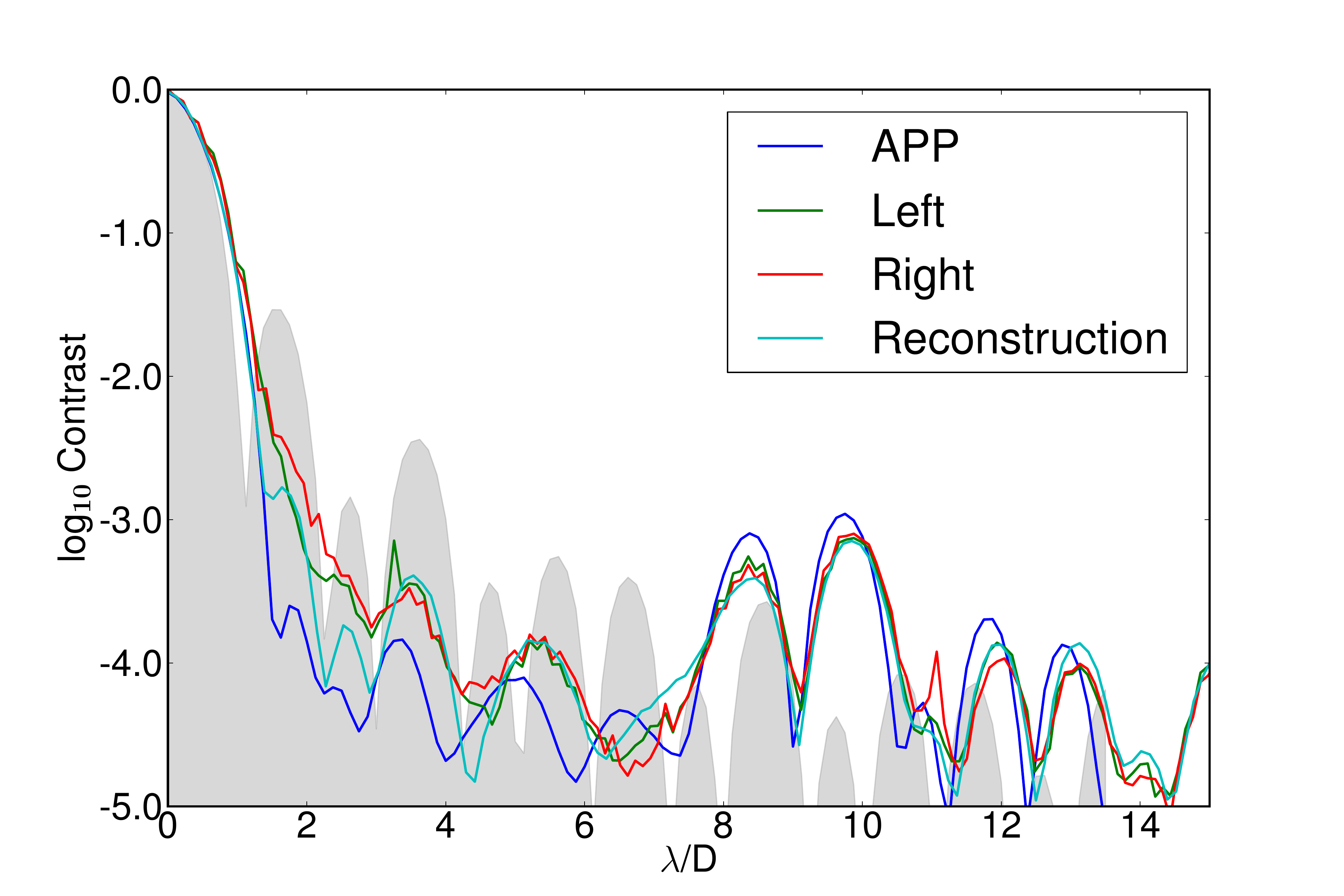}}
\caption{Contrast curves for the two PSFs of the vAPP as measured in the lab at 750 nm compared to the ideal case, and compared to a reconstruction that used the pupil measurements as input. The lab measurements are shown in green and red for respectively the Left and Right PSFs. The theoretical contrast is shown in dark blue, and the contrast of the model PSF using the reconstructed transmission, fast axis and retardance is plotted in light blue. The gray shaded area shows the contrast for an unaberrated PSF.} 
\label{fig:psf_recon_750}
\end{figure}

In order to measure the PSF of the vAPP we use the laboratory setup shown in Fig. \ref{fig:schematic_psf}. 
The optics from the fiber to the vAPP are identical to the pupil imaging setup in Fig. \ref{fig:pup_setup}.
As a point source we use a Thorlabs SM600 fiber that is single mode between 550 and 800 nm and has a mode field diameter that is significantly smaller than the optical resolution of the setup.
This point source is re-imaged onto the pupil using a $2''$ achromatic lens Thorlabs $f=250\,\mathrm{mm}$ lens.
Behind the vAPP, a combination of an ``achromatic'' (i.e., Quartz/MgF$_2$) Thorlabs quarter-wave plate (AQWP05M-600) and a Wollaston prism splits the two circular polarization states into two beams angularly separated by 0.5 degrees. 
A $f=250\,\mathrm{mm}$ lens forms two PSF images on the CCD camera (SBIG ST2000XM). 
This optical setup is similar to the way a vAPP would be implemented in a science camera on a telescope.
With approximately 500 mm between the Wollaston prism and the CCD, the separation of the two complementary PSFs is 4.36 mm. 
Both PSFs are simultaneously imaged on the CCD. 
At 400 nanometers the FWHM of the PSF is about 22 microns, and with the 7.4 micron pixel size this means the PSF is always Nyquist-Shannon sampled with at least 3 pixels per diffraction element.
We require images with high dynamic contrast of at least $10^{-4}$ to characterize the vAPP PSFs.
The dynamic range of the camera is limited to approximately $2.5 \times 10^{-3}$. 
We therefore combine images with different exposure times and construct a high dynamic range (HDR) image.
Images are taken at five different exposure times (i.e., 10, 100 milliseconds, 1, 10 and 30 seconds). 
For each exposure time 10 images are taken and median combined after subtracting the master bias.
To obtain HDR images we start with the median combined frame that has the shortest exposure time (i.e., 10 milliseconds) and replace the pixels of this image with exposure-time corrected values of the pixels that are in the linear regime of the next longest exposed median combined image. 
This procedure is repeated iteratively with increasingly longer integrations. 
After assembling these HDR images, the Left beam and Right beam PSFs are extracted by using a square box of 140 by 140 pixels centered on the pixel with the highest value in the core of each PSF. 
The background is estimated by taking the mean value of a rectangular region (100 by 200 pixels) located between the two PSFs and subtracted from both PSF images. 
Even at the longest wavelength of 800 nm this background region is located at $\sim 25 \lambda/D$ from the central core of both of the PSFs and the contribution of the PSF halo to the flux in this region is negligible. 
Each individual PSF is then normalized by its peak flux. 
These measurements are repeated at wavelengths from 500 to 800 nanometers at 50 nm intervals with 10 nm wide filters. 
Figure~\ref{fig:allpsfs} presents the Left and Right beams of the PSFs at increasing wavelengths.
The coronagraph shows consistent performance across all wavelengths, which demonstrates its broadband performance. 
The contrast inside the dark hole barely varies with wavelength, and therefore the dominating chromatic effect is the fundamental growth of the diffraction pattern with wavelength.
This means that for imaging through broadband filters, the inner working angle is only limited by the longest transmitted wavelength.
Figure~\ref{fig:psf_recon_750} presents the contrast curve for 750 nm by taking azimuthal averages of normalized Left and Right beam PSFs. 
The flux at a given radius $r$ is calculated by taking the mean over a region defined between radius $r$ and $r+\Delta r$ and an angular separation 135 degrees wide in the dark hole.

\begin{figure}[!ht]
\centerline{\includegraphics[scale=.35]{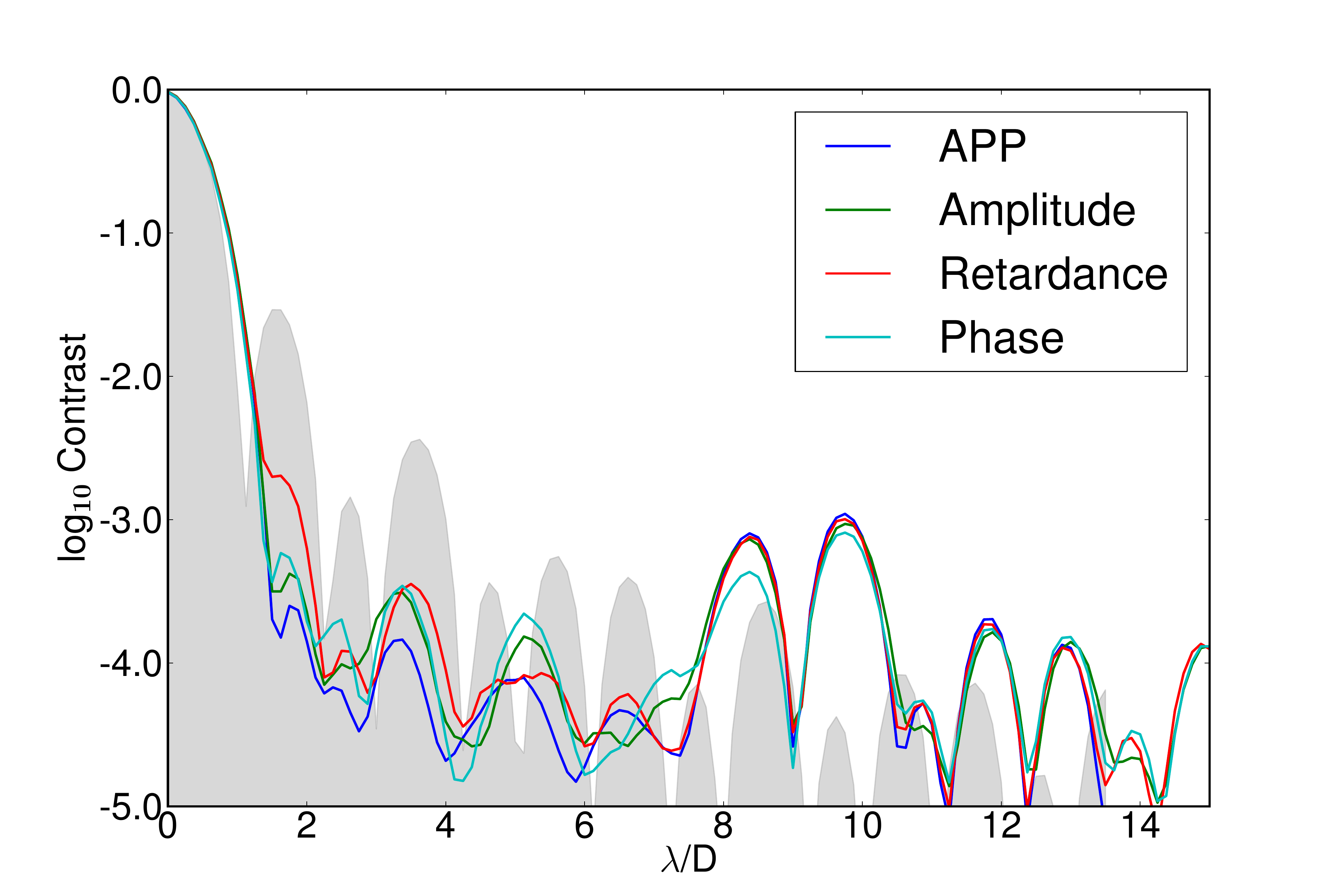}}
\caption{Theoretical contrast curves of the most important influences on the vAPP's contrast degradation. Using the model described in Section \ref{sec:3} the theoretical PSF was created using measured offsets to only one of the three properties (i.e., transmission, fast axis or retardance) while keeping the other two properties at their ideal values. The theoretical APP curve is shown in blue. The gray shaded area shows the contrast for an unaberrated PSF.}
\label{fig:contrast2}
\end{figure}

\begin{figure}[!h]
\centerline{\includegraphics[scale=.35]{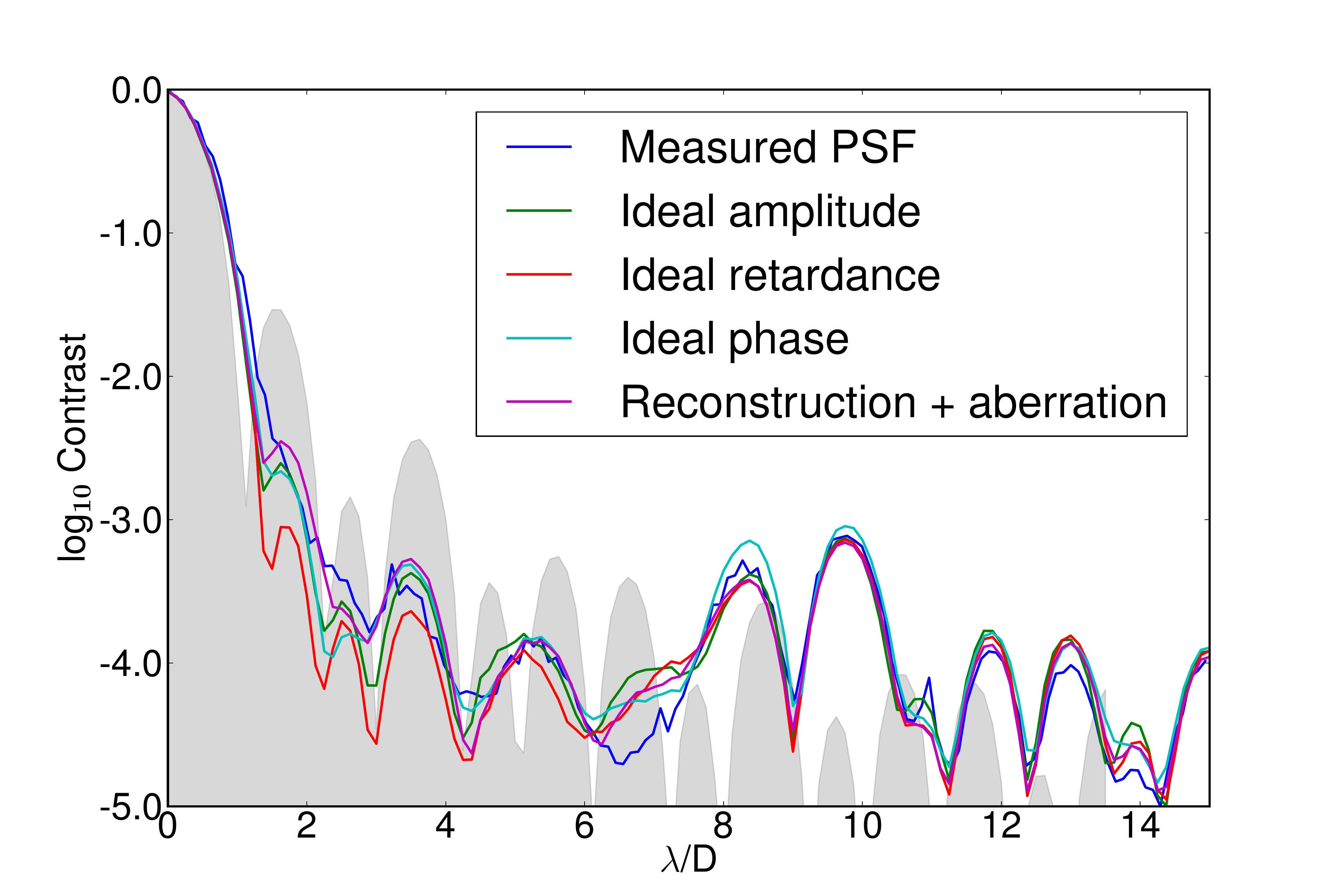}}
\caption{Theoretical contrast curves of the most important combination of influences on the vAPP's contrast. Using the model described in Section \ref{sec:3} the theoretical PSF is created using two of the three measured properties (i.e., transmission, fast axis or retardance) while keeping the other property at its ideal value. The dark blue line shows the average of the two measured PSFs. The purple line shows the reconstruction using all three measured quantities together with the QWP performance and 0.15 radians of spherical aberration. The shaded gray area shows the contrast for an unaberrated PSF.}
\label{fig:contrast3}
\end{figure}

In order to model the PSFs of the vAPP, we require a formalism that can account for relative phase in the pupil plane so that we can calculate the intensity (i.e., the square of the amplitude) in the final focal plane using Fraunhofer diffraction theory. 
The Mueller-Stokes theory cannot account for this, so we use the Jones formalism instead, as the split beams are 100\% polarized. 
The PSF model is detailed in Appendix \ref{sec:appb}.
Using the measured phase pattern, retardance and transmission of the plate, the QWP retardance $\delta_\mathrm{QWP}$ from the Thorlabs specifications, and the model described in Appendix \ref{sec:appb} we reconstruct the expected PSFs. 
The reconstructed PSFs are normalized to their peak flux to enable comparison with lab measurements of the PSFs in Figs.~\ref{fig:psf_recon_750}, \ref{fig:contrast2} and \ref{fig:contrast3}.

Figure~\ref{fig:psf_recon_750} shows the contrast profile of Left and Right beam PSFs, which are extremely similar.
The measurements show a contrast degradation of typically 0.5 decades with respect to the ideal curve. 
Inside the dark hole the contrast curve reconstruction agrees with the measured Left and Right beam PSFs at all angular distances, apart from $2\ \mathrm{and}\ 7\ \lambda/D$. 
To establish which of the leakage terms and other degrading terms is most detrimental we consider in Fig.~\ref{fig:contrast2} the effect of the individual properties on the contrast, and model the PSF for each property separately while keeping the other two theoretically perfect. 
This demonstrates that, by itself, the retardance has the largest impact on the contrast in the dark hole of the prototype, as it immediately leads to leakage of the original PSF.
We also calculate the impact of improving only one of the three proprties that we have characterized (see Fig. \ref{fig:contrast3}). 
Again our conclusion is that improving the retardance in subsequent plate manufacture or filtering leakage terms \cite{Mawet:09,Snik:14spie} will improve the average contrast by 0.35 decades between 2 and 7 $\lambda/D$ towards the ideal.

A residual discrepancy between measurements and reconstruction is observed in Fig.~\ref{fig:psf_recon_750}.
We attribute these differences at these angular separations to low-order wavefront aberrations in the re-imaging optics, variations in thickness of the vAPP substrate and the Wollaston prism.
In Figure~\ref{fig:contrast3} we show that injecting 0.15 radians RMS of third-order spherical aberrations results in excellent agreement of the model with the measured curve at these angular scales. 
Such wavefront aberrations can be corrected in a telescope (for at least one PSF) by offsetting the deformable mirror. 
\FloatBarrier
\section{Conclusions \& Outlook}
\label{sec:5}
In this paper we present the characterization of a prototype of the vector Apodizing Phase Plate Coronagraph.
Owing to the application of the vector phase applied by patterned Multi-Twist Retarder liquid crystals, the coronagraphic performance is achieved over an unprecedented wavelength range of 500--900 nm in two complementary PSFs.
With lab measurements we characterize the phase pattern, the retardance and the transmission of the constructed vAPP prototype, and measured and modeled the resultant PSFs. 
We compared the measured PSFs with the modeled PSFs and find excellent agreement at all wavelengths and angular separations when all known non-ideal effects are included.
The vAPP provides up to two decades of suppression at 2 $\lambda/D$ and keeps the diffraction halo at a level of $10^{-3.8}$ of the peak intensity out to 7 $\lambda/D$ as measured in a 135 degree wedge
The retardance measurements show a wavelength behavior that is consistent with a 3-layer MTR, which provides coronagraphic performance over almost an order of magnitude wider in spectral bandwidth than the previous APP coronagraph.
Still, the most significant degradation from the ideal APP PSF is due to offsets from half-wave retardance. 
Subsequent iterations of vAPP coronagraphs should therefore focus on improving the manufacturing technique with respect to the retardance and/or filter the leakage terms.

\begin{figure}[!htb]
\centerline{\includegraphics[scale=.35]{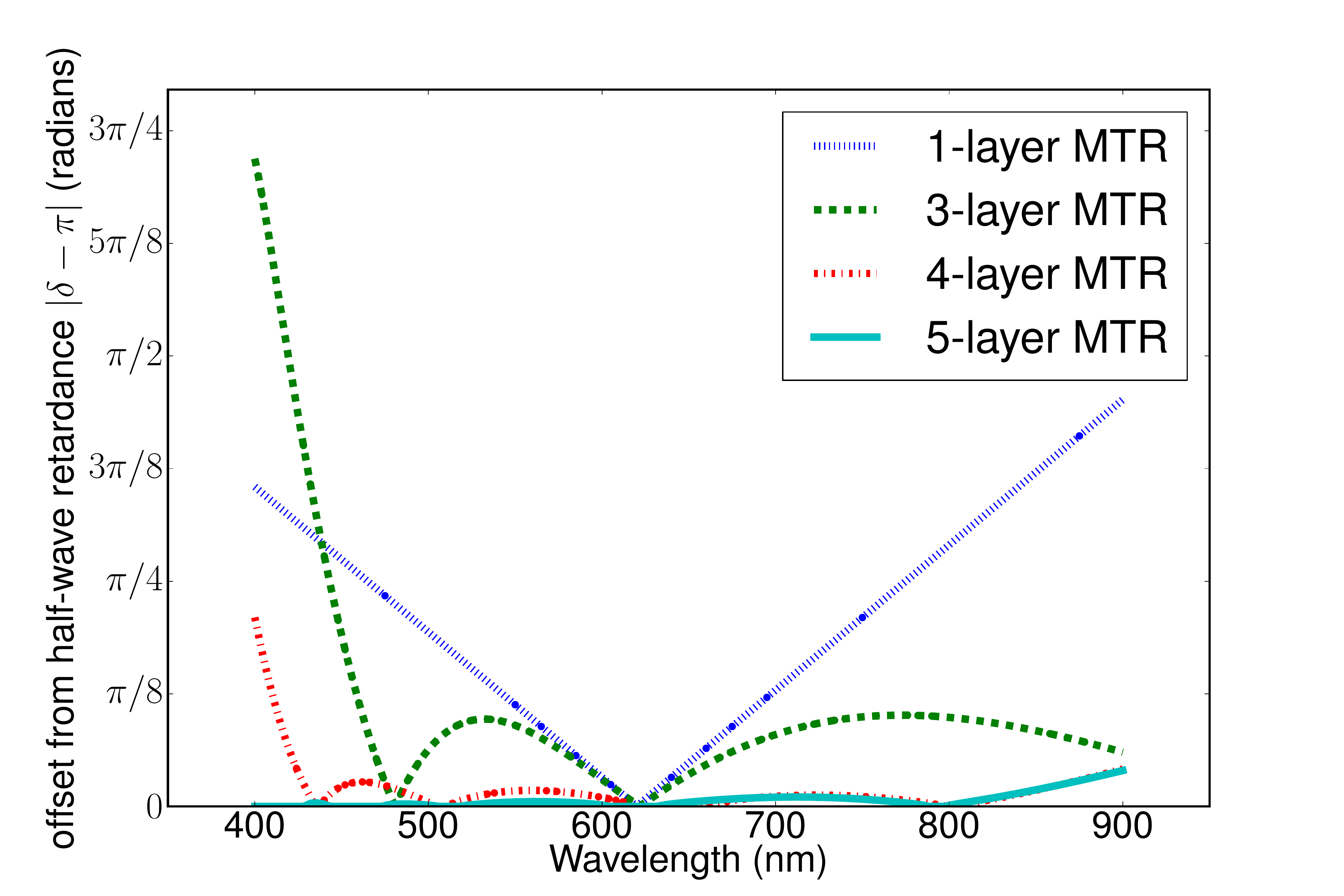}}
\caption{Theoretical retardance curves as a function of wavelength for different layered MTR designs. This paper describes the performance of a 3-layer MTR device. Increasing the number of twisting layers in the MTR improves the retardance performance across the wavelength range.}
\label{fig:mtr_retardance}
\end{figure}

\begin{figure}[!htb]
\centerline{\includegraphics[scale=.35]{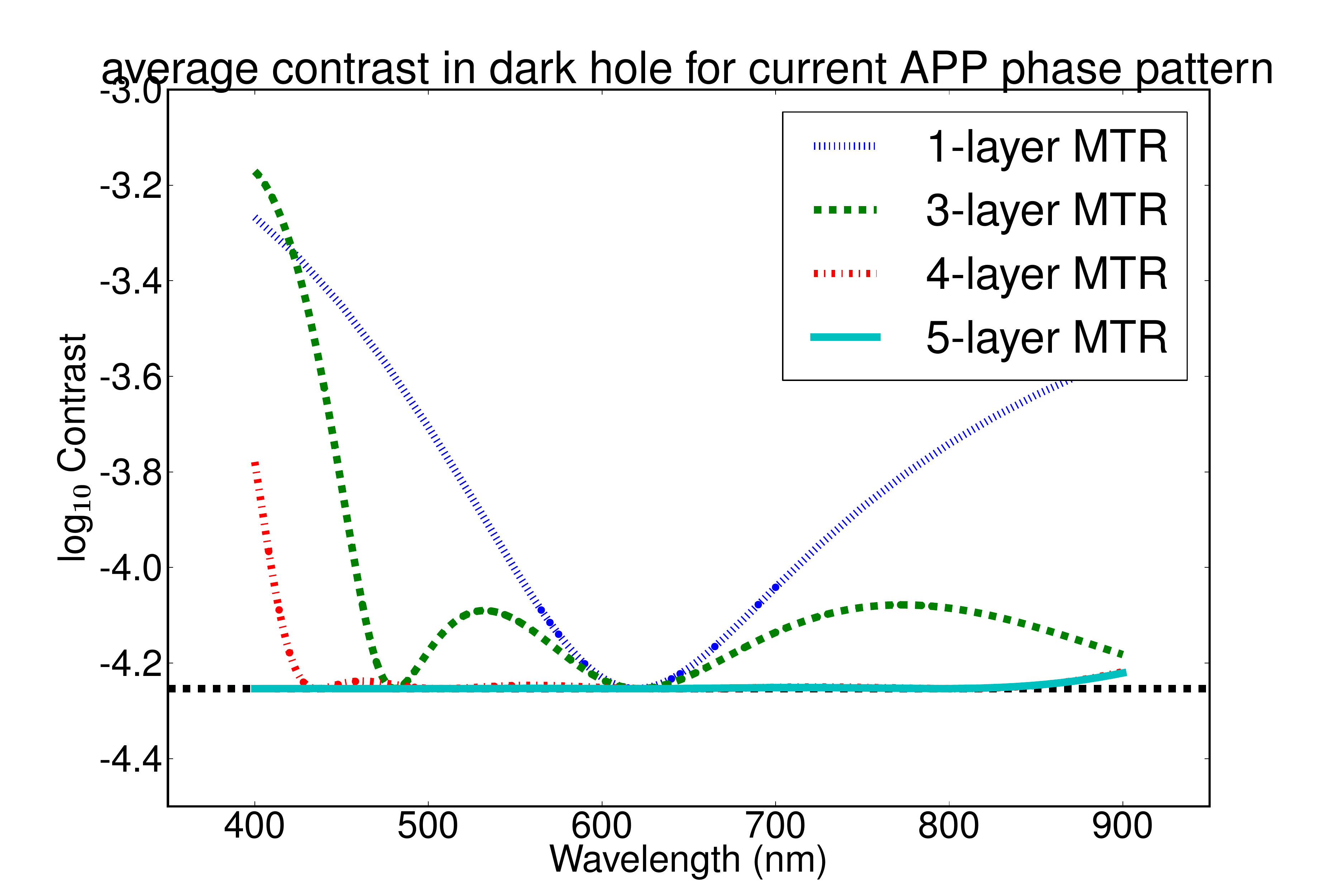}}
\caption{Theoretical contrast curves as a function of wavelength for different layered MTRs and the current vAPP phase pattern design. The contrast was derived from Fig. \ref{fig:mtr_retardance} by creating PSF models and measuring the depth of the dark hole between 2 and 7 $\lambda/D$ in a 135 degree wedge. This paper describes the performance of a 3-layer vAPP device. The black dashed line shows the theoretical perfect contrast given the adopted phase pattern for a perfect half-wave retarder across all wavelengths. Even though the contrast seems to bottom out for device with more than three layers, the contrast could be improved by adopting a more extreme phase pattern with higher contrast.}
\label{fig:mtr_contrast}
\end{figure}

The retardance as a function of wavelength could still be further improved by adding more liquid crystal layers. 
For instance, a 4-layer device has significantly better  performance than the 3-layer design that was used in this research. 
The improvement to the theoretical retardance and contrast are shown in Fig. \ref{fig:mtr_retardance} and Fig. \ref{fig:mtr_contrast} respectively. 

The contrast in Fig. \ref{fig:mtr_contrast} bottoms out as it reaches the limit of the adopted APP phase pattern. 
Owing to the direct-write technique, the vAPP permits the adoption of more extreme phase patterns that reach much deeper (theoretical) contrasts (e.g.~ \cite{Carlotti:13b,Otten:14spie}).
Using the 200 by 200 pixels of this prototype it can be calculated that the outer working angle can theoretically go up to 100 $\lambda/D$.
The production of such an upgraded vAPP coronagraph will be subject to more demanding production tolerances.
Particularly the retardance offsets will need to be much smaller than for the current prototype, or the spectral bandwidth will need to be decreased.
However, the leakage terms can also be avoided by applying circular polarization filtering on either side of the vAPP \cite{Snik:14spie}.
An alternative solution is found by integration of the circular polarization beam-splitting in the vAPP, by using a polarization grating \cite{Otten:14spie}.
In this case, all leakage terms end up in a third, unaberrated PSF.

More extreme vAPP devices will also be more demanding in terms of the phase pattern itself.
The accuracy of the fast axis orientation (and hence the vector phase) pattern depends on the accuracy of the angle of linear polarization of the UV laser that is used for the photo-alignment layer. 
The latest manufacturing setup has a Pockels cell to better control this angle and dramatically increases the speed at which the phase patterns are realized. 
Furthermore, the instrumental polarization issues due to the fold mirror should be avoided.

The measurement setup that we use for the reconstruction of the vAPP device properties is relatively easy to implement and use, but it has several limitations. 
The accuracy of the reconstruction is dominated by the photon noise on the input images. 
Unfortunately, the small fiber and narrowband filters attenuate light and therefore require long exposure times. 
The light source exhibits fluctuations especially within the first few hours after being switched on, which influence the accuracy of the reconstruction.
The reconstruction method removes first order fluctuations of the light but is sensitive to fluctuations between measurements at orthogonal angles of the polarizer.
By using a brighter light source, a larger diameter fiber, and/or a camera with better quantum efficiency we can minimize these effects by reducing measurement noise and taking the orthogonal exposures with shorter time intervals.
Another solution for the fluctuations would be to independently monitor the light flux out of the fiber to allow for normalization in data reduction.

While the coronagraph in this paper is designed to work at visible wavelengths, most high-contrast imaging instruments typically work at near-infrared wavelengths. 
Fortunately the wavelength range at which the plate optimally works can be optimized by a judicial choice of different liquid crystals and substrates for the vAPP. 
Using these methods \cite{Packham:10} have demonstrated a polarization grating system that works from 3 to 20 microns, and showed that the retardance is only marginally impacted (3\%) at cryogenic temperatures. If necessary, such a change in retardance can be pre-compensated during manufacturing.

Instruments that would be suitable for a vAPP coronagraph are adaptive optics-fed high-contrast cameras such as SPHERE-ZIMPOL \cite{Roelfsema:10} on the Very Large Telescope, GPI on Gemini South \cite{Macintosh:14}, ExPo on the William Herschel Telescope \cite{Rodenhuis:12}, and MMT-Pol \cite{Packham:12}. 
These polarimetric instruments already contain the required Wollaston prism (or other polarizing beam-splitter) to split the PSFs. 
Note that even in cases of moderate ($\sim$50\%) Strehl ratio, the vAPP still delivers a average raw contrast of $10^{-3.9}$ at 4--6 $\lambda/D$.
The second vAPP prototype (cf.~the design presented by \cite{Otten:14spie}) has recently been integrated in LMIRCam on the Large Binocular Telescope \cite{Skrutskie:10}.

Moreover, as the vAPP coronagraph naturally pairs with polarimetry it is possible to reach an even deeper contrast for scattered planet light by combining the two techniques \cite{Snik:14spie}. 
The two complementary images may also furnish focal-plane wavefront sensing cf.~\cite{Riaud:12} to provide information for real-time or post-factor correction of (instrumental) wavefront aberrations.
Finally, we must emphasize that the use of the achromatic patterning technique is not limited to phase manipulation in pupil-plane coronagraphs, but can also be used for focal-plane coronagraphs or a combination of both. 

\section*{Acknowledgments}
The authors would like to thank Remko Stuik and Gerard van Harten for their helpful suggestions on the laboratory measurements. Furthermore, we thank the anonymous referees for their helpful suggestions which improved this manuscript. This work is part of the research programme Instrumentation for the E-ELT, which is partly financed by the Netherlands Organisation for
Scientific Research (NWO).

\appendix
\section{Pupil measurement theory and setup}
\label{sec:appa}
\subsection{Pupil reconstruction}
\label{sec:appa1}
This subsection provides the conventions and theory to describe the measurements of the vAPP properties with the setup presented in Fig.~\ref{fig:pup_setup}, and the subsequent data reduction.

The partial polarization of beam of light can be described by a 4-element Stokes vector. 
The four elements are $I$, $Q$, $U$ and $V$ where $I$ denotes intensity, $Q$ and $U$ denote linear polarization at $0/90$ degrees and $\pm45$ degrees, respectively, and $V$ denotes circular polarization. 
A $4\times4$ Mueller matrix describes polarization-changing optical components or media that convert an input Stokes vector into a new Stokes vector. 
Mathematically, the output Stokes vector can be calculated by multiplying the Mueller matrix against the Stokes input vector as seen in Eq. \eqref{eq:mm}.
\begin{equation}
\mathbf{S_\mathrm{out}}=\mathbf{M}\,\mathbf{S_\mathrm{in}}=\left( \begin{matrix} 
m_{00} & m_{01} & m_{02}& m_{03} \\
m_{10} & m_{11} & m_{12}& m_{13}\\
m_{20} & m_{21} & m_{22}& m_{23}\\
m_{30} & m_{31} & m_{32}& m_{33}
\end{matrix} \right)\,\left( \begin{matrix} 
I_\mathrm{in}\\
Q_\mathrm{in}\\
U_\mathrm{in}\\
V_\mathrm{in}
\end{matrix} \right)
\label{eq:mm}
\end{equation}

An ideal retarder with a certain retardance $\delta$  and a fast axis of $\theta=0$ radians has the following Mueller matrix:
\begin{equation}
\mathbf{M_\mathrm{HWP,0}}(\delta)=\left( \begin{matrix} 
1 &0& 0& 0 \\
0&1&0&0\\
0&0 & \cos{\delta}&\sin{\delta} \\
0&0& -\sin{\delta}&\cos{\delta}
\end{matrix} \right).\end{equation}
As the vAPP locally has a certain fast axis $\theta = \frac{\phi_\mathrm{APP}}{2}$ we rotate this Mueller matrix with the rotation matrix $\mathbf{T}_\mathrm{M}(\theta)$:
\begin{equation}
\mathbf{T}_\mathrm{M}(\theta)=\left( \begin{matrix} 
1 &0& 0& 0 \\
0&\cos{2\theta} & \sin{2\theta}&0\\
0&-\sin{2\theta}& \cos{2\theta}&0\\
0&0&0&1
\end{matrix} \right)\end{equation} creating the equation \begin{equation}
\mathbf{M_\mathrm{HWP,\theta}}(\delta)=\mathbf{T}_\mathrm{M}(-\theta)\,\mathbf{M_\mathrm{HWP,0}}(\delta)\,\mathbf{T}_\mathrm{M}(\theta).
\end{equation}
Since this retarder is positioned between 2 polarizers with a certain position angle $\alpha$ with respect to the anti-symmetry axis we describe each polarizer with the following Mueller matrix
\begin{equation}
\mathbf{M_\mathrm{pol,\alpha}}=\mathbf{T}_\mathrm{M}(-\alpha)\,\frac{1}{2}\left( \begin{matrix} 
1 &1& 0& 0 \\
1&1 & 0&0 \\
0&0& 0&0\\
0&0&0&0
\end{matrix} \right)\,\mathbf{T}_\mathrm{M}(\alpha).
\end{equation}
\noindent In the case of parallel polarizers we obtain the equation
\begin{equation}
\mathbf{M_\mathrm{setup,\parallel}}(\alpha,\delta,\theta)=\mathbf{M_\mathrm{pol,\alpha}}\,\mathbf{M_\mathrm{HWP,\theta}}(\delta)\,\mathbf{M_\mathrm{pol,\alpha}}.
\end{equation}
\noindent With crossed polarizers the Mueller matrix for the whole setup becomes
\begin{equation}
\mathbf{M_\mathrm{setup,\perp}}(\alpha,\delta,\theta)=\mathbf{M_\mathrm{pol,\alpha+\pi/2}}\,\mathbf{M_\mathrm{HWP,\theta}}(\delta)\,\mathbf{M_\mathrm{pol,\alpha}}.
\end{equation}
\noindent The input light can be polarized or unpolarized with an intensity $I_\mathrm{in}$ and is represented by the Stokes vector
\begin{equation}
\mathbf{S_\mathrm{in}}=\left( \begin{matrix} 
I_\mathrm{in} \\
Q_\mathrm{in} \\
U_\mathrm{in}\\
V_\mathrm{in}
\end{matrix} \right).
\end{equation}
The output vector then becomes
\begin{equation}
\mathbf{S_\mathrm{out}}=\mathbf{M_\mathrm{setup,\parallel/\perp}}(\alpha,\delta,\theta)\,\mathbf{S_\mathrm{in}}.
\label{eq:sout}
\end{equation}

Since we only measure the intensity of the vAPP plate in between polarizers we only look at the first element of the output Stokes vector $S_\mathrm{out}$. 
This element will be referred to as $I_{\alpha,\parallel/\perp}(\delta,\theta)$. 
As an example; for the intensity of the plate between crossed polarizers with a position angle of 45 degrees the intensity is $I_{45^\circ,\perp}(\delta,\theta)$. 
A combination of intensity ratios gives us an equation with no dependency on the fast axis, input intensity or polarization, or the transmission:
\begin{equation}
\frac{I_{0^\circ,\parallel}-I_{0^\circ,\perp}}{I_{0^\circ,\parallel}+I_{0^\circ,\perp}}+\frac{I_{45^\circ,\parallel}-I_{45^\circ,\perp}}{I_{45^\circ,\parallel}+I_{45^\circ,\perp}}=1+\cos(\delta).
\label{eq:retardance0}
\end{equation}
Similarly, the equation
\begin{equation}
\frac{I_{0^\circ,\parallel}-I_{0^\circ,\perp}}{I_{0^\circ,\parallel}+I_{0^\circ,\perp}}-\frac{I_{45^\circ,\parallel}-I_{45^\circ,\perp}}{I_{45^\circ,\parallel}+I_{45^\circ,\perp}}=2 \cos(4 \theta) \sin(\delta/2)^2
\label{eq:fastaxis0}
\end{equation}
gives us an expression that yields the retarder orientation (i.e.~half the vector phase) that is insensitive to the input intensity and polarization as long as they are constant during the exposures.

The unknown retardance $\delta$ can be reconstructed with the equation
\begin{equation}
\delta=\arccos\left(\frac{I_{0^\circ,\parallel}-I_{0^\circ,\perp}}{I_{0^\circ,\parallel}+I_{0^\circ,\perp}}+\frac{I_{45^\circ,\parallel}-I_{45^\circ,\perp}}{I_{45^\circ,\parallel}+I_{45^\circ,\perp}}-1\right)
\label{eq:retardance}
\end{equation}
which is the inverse of Eq. \eqref{eq:retardance0}.
\noindent Using this retardance, the fast axis $\theta$ can be calculated using a rewritten version of Eq. \eqref{eq:fastaxis0}
\begin{equation}
\theta=\frac{1}{4} \arccos\left(\frac{1}{2} \frac{\frac{I_{0^\circ,\parallel}-I_{0^\circ,\perp}}{I_{0^\circ,\parallel}+I_{0^\circ,\perp}}-\frac{I_{45^\circ,\parallel}-I_{45^\circ,\perp}}{I_{45^\circ,\parallel}+I_{45^\circ,\perp}}}{\sin(\delta/2)^2}\right)
\label{eq:fastaxis}
\end{equation}

Note that this equation will produce fast axis values that fold at $\pi/4$, because of the properties of the arccosine function. 
This problem will be resolved at the end of the following Subsection.

As a consistency check, the retardance was also estimated by looking at the flux ratio
\begin{equation}
\frac{I_{45^\circ,\parallel}}{I_{45^\circ,\perp}+I_{45^\circ,\parallel}}
\end{equation}
in the anti-symmetry axis of the vAPP pattern, and assuming that the polarizers are at 45$^\circ$ with respect to the fast axis at that location. 
If the plate is exactly half-wave, the second polarizer will block the incoming light.
A non-zero flux ratio indicates an offset from half-wave.
\begin{equation}
\delta=\arccos\left(2 \, (I_{45^\circ,\parallel}/(I_{45^\circ,\perp}+I_{45^\circ,\parallel}))-1\right).
\label{eq:methodb}
\end{equation}
relates this flux ratio to a retardance estimate. 

We call reconstructing the retardance using Eq. \eqref{eq:fastaxis} Method A and reconstructing using Eq. \eqref{eq:methodb} Method B. 
To record these intensity measurements we construct a laboratory setup to measure the intensity across the vAPP clear aperture as seen through polarizers pairs. 
In total we need 8 different polarizer pair positions and corresponding intensity measurements. 
The list is seen in Tab.~\ref{tab:pols} and Fig.~\ref{fig:polarizers}.

\begin{table}[h]
  \caption{List of used polarizer positions. An example of the corresponding measured intensities can be seen in Fig. \ref{fig:polarizers}. }
  \begin{center}
    \begin{tabular}{ccc}
    \hline
    Polarizer 1 & Polarizer 2 & Intensity\\
		\hline
    $0\,^{\circ}$ & $0\,^{\circ}$ & $I_{0^{\circ},\parallel}$\\
    $0\,^{\circ}$ & $90\,^{\circ}$ & $I_{0^{\circ},\perp}$\\
    $45\,^{\circ}$ & $45\,^{\circ}$ &$I_{45^{\circ},\parallel}$\\
		$45\,^{\circ}$ & $135\,^{\circ}$ & $I_{45^{\circ},\perp}$\\
		$22.5\,^{\circ}$ & $22.5\,^{\circ}$ &$I_{22.5^{\circ},\parallel}$\\
		$22.5\,^{\circ}$ & $112.5\,^{\circ}$ &$I_{22.5^{\circ},\perp} $\\
		$67.5\,^{\circ}$ & $67.5\,^{\circ}$ & $I_{67.5^{\circ},\parallel}$\\
		$67.5\,^{\circ}$ & $157.5\,^{\circ}$ & $I_{67.5^{\circ},\perp}$\\
		\hline
    \end{tabular}
  \end{center}
	\label{tab:pols}
\end{table}

\subsection{Pupil measurement setup \& data reduction}
\label{sec:labsetup}
\label{sec:appa2}
The pupil measurement setup consists of two polarizers and optics that re-image the vAPP pupil onto the SBIG ST2000XM CCD. 
A schematic overview of the layout is presented in Fig.~\ref{fig:pup_setup}. 
A Xenon arc lamp source from Cairn Research is focused into a single-mode SM600 fiber specified to be single mode between 550 and 800 nanometers. 
The fiber has a mode field diameter of $4.3\ \mu m$ at 633 nm which is substantially smaller than the optical resolution of our setup.
Thorlabs FKB-VIS-10 10 nm narrowband interference filters are placed between the Xenon source and the fiber to select quasi-monochromatic light (i.e., 500-800 nanometers at 50 nm steps).
A $f=250\,\mathrm{mm}$ lens re-images the fiber to form a pupil on the APP. 
This 5.5 mm pupil of the vAPP is then re-imaged onto the 1600 by 1200 pixel detector with 7.4 micron pixel size. 
A combination of a $f=250\,\mathrm{mm}$ and $f=300\,\mathrm{mm}$ lens form a re-imaged pupil size of 6.6 mm on the 11.8 by 8.9 mm chip. 

\begin{figure}[h]
\centerline{\includegraphics[scale=0.85]{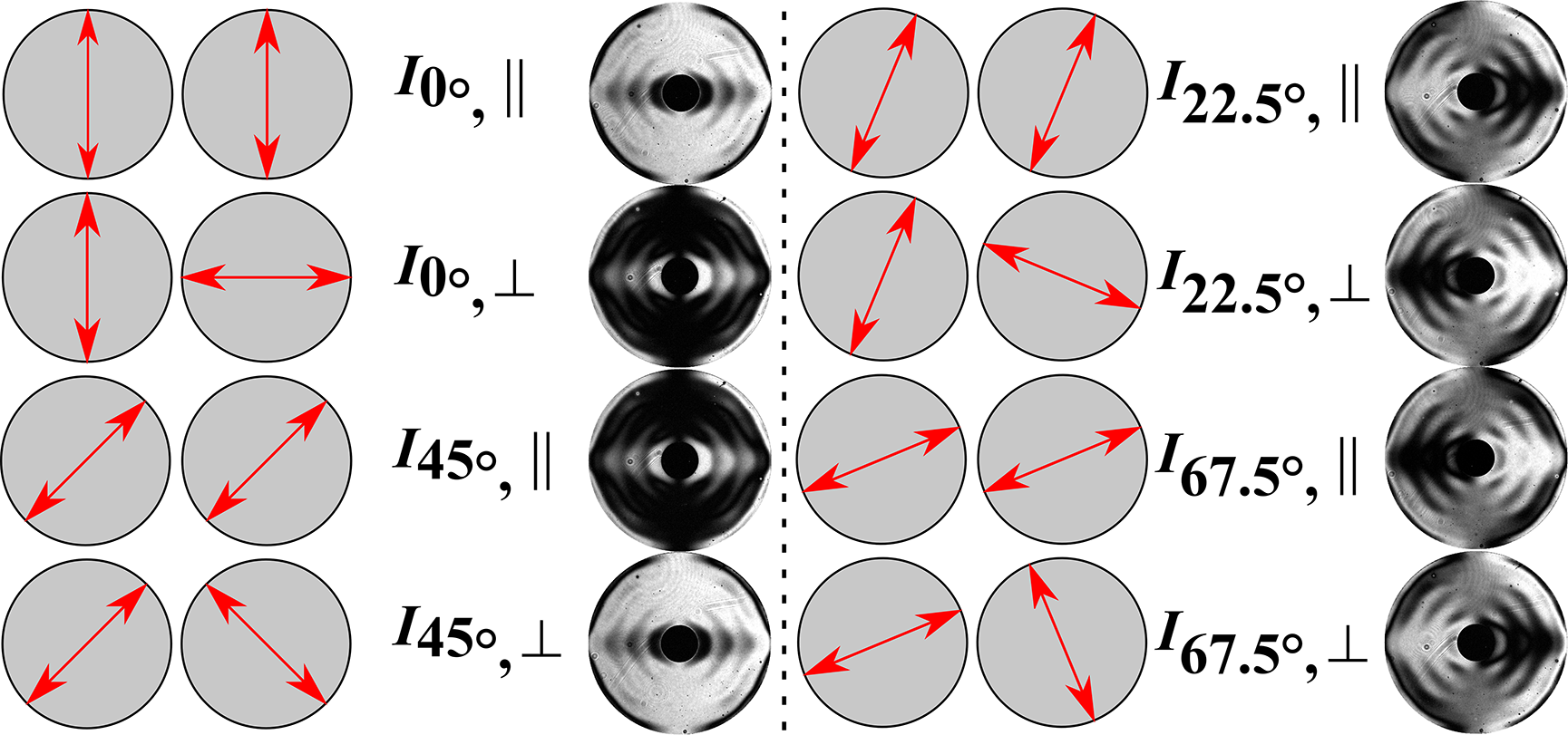}}
\caption{Polarizer orientation angles and resultant measured intensity map for each set of polarizer orientations. Using these 8 measurements we can reconstruct the transmission, fast axis and retardance of the optic.}
\label{fig:polarizers}
\end{figure}

The measurements are recorded at the defined polarizer angles and at wavelengths from 400 to 800 nanometers at 50 nm intervals. 
The different angles used for the reconstruction are listed in Table \ref{tab:pols} and shown in Fig. \ref{fig:polarizers} together with the measured intensity images.
The values in Fig. \ref{fig:fastaxisoffset} are used to offset the angles of the polarizer pair between wavelengths. 
The offsets are fine-tuned by looking at the intensity image $I_{0^\circ,\parallel}$ and optimizing the symmetry of this image.

The exposure time of the individual frames for these pupil measurements is 100 seconds because of the narrow spectral band and single-mode fiber. 
At each polarizer pair orientation, one exposure followed by one bias frame is repeated 10 times to form a total integration time of 1000 seconds. 
Additionally we also take 20 dark frames at 100 seconds each to form a master dark.
The bias frames are median combined to create a master bias and similarly after subtracting the master bias from each individual frame the dark frames are median combined into a master dark. 
After removing the bias and dark from all the frames, they are stacked to form a final image. 
This is repeated for each wavelength and polarizer pair image. 

To estimate the background in scattered light along with low spatial frequency structure, we use the region of the CCD that corresponds to a region that is blocked by the amplitude mask of the vAPP. 
After masking out the APP pupil image and determining the center of the pupil image, we fit a function of the form $f(x,y)=a\, x+b\, y+c+d\,(x^2+y^2)$ and we solve for $a,b,c\ \mathrm{and}\ d$ using a least-square approach, and subtract it off the image.

Using Eq. \eqref{eq:retardance} we reconstruct the retardance from two sets of four corrected images.  
The arccosine in the equation is only defined for values between -1 and 1, and therefore the factor $\frac{I_{0^\circ,\parallel}-I_{0^\circ,\perp}}{I_{0^\circ,\parallel}+I_{0^\circ,\perp}}+\frac{I_{45^\circ,\parallel}-I_{45^\circ,\perp}}{I_{45^\circ,\parallel}+I_{45^\circ,\perp}}-1$  is susceptible to noise. 
The factors that are outside of these limits are truncated to the nearest defined value of either -1 or 1 before taking the arccosine. 
Similarly, when computing the fast axis using Eq. \eqref{eq:fastaxis} this procedure is used to ensure that all values are real.

The reconstructed fast axis values reflect at $\pi/2$ due to the invariance of the polarizer under rotations of 180 degrees. 
If the fast axis orientation is larger than $\pi/2$ radians the intensity decreases again creating a non-unique relation between the intensity ratio and phase.
To overcome this we obtain additional constraints by repeating these measurements with an offset of 22.5 degrees. 
The intensity and corresponding reconstructed phases fold back at different locations in the images while the retardance remains the same as before. 
The input phase pattern is used together with the two reconstructed fast axis maps in order to determine the fast axis orientation.
With more signal to noise per pixel it would be possible to directly unfold the reconstructed fast axis using the input phase map by snapping to the phase solution that matches the input most closely.
In our case the individual images have too much noise to directly convert using the known phase pattern. Each of the image pairs that are taken have areas where the fast axis is linear as a function of the input phase and does not fold backwards. The linear areas of the images are stitched together. At locations where $\left|{\phi}\right|>\pi/2$ radians a degeneracy still exists. This is solved by applying the appropriate sign and offset to this area of the reconstructed fast axis.

\subsection{Error propagation}
\label{sec:appa3}
To determine the effect of misaligned polarizers and measurement noise on the reconstructed retardance and fast axis, we simulate intensity patterns using Eq. \eqref{eq:sout} and introduce small offsets to the polarizers ($<$2 degrees) and simulated photon shot noise by injecting Poisson noise into these images. 
We then reconstruct unwrapped fast axis and retardance maps using the aforementioned method.
From these simulations we see that identical angular offsets to the polarizer pair positions have no significant effect on the retardance reconstruction. 
Offsetting the polarizer pairs does change the reconstruction of the fast axis pattern. 
This polarizer offset slightly shifts the position at which the phase folds in individual images.
This influences the reconstructed (un-folded) phase maps that joins these individual images although its impact depends how far the polarizers are offset.
The relative positions between the polarizers can be determined to within 5 arcminutes, but there is an unknown but fixed offset (on the order of 1 degree) between the polarizer mount position and the fast axis orientation of the APP. 
The simulations show that offsetting the polarizers by 1 degree does not significantly affect the reconstruction of the phase apart from a piston term.

\begin{figure}[h]
\centerline{\includegraphics[scale=.3]{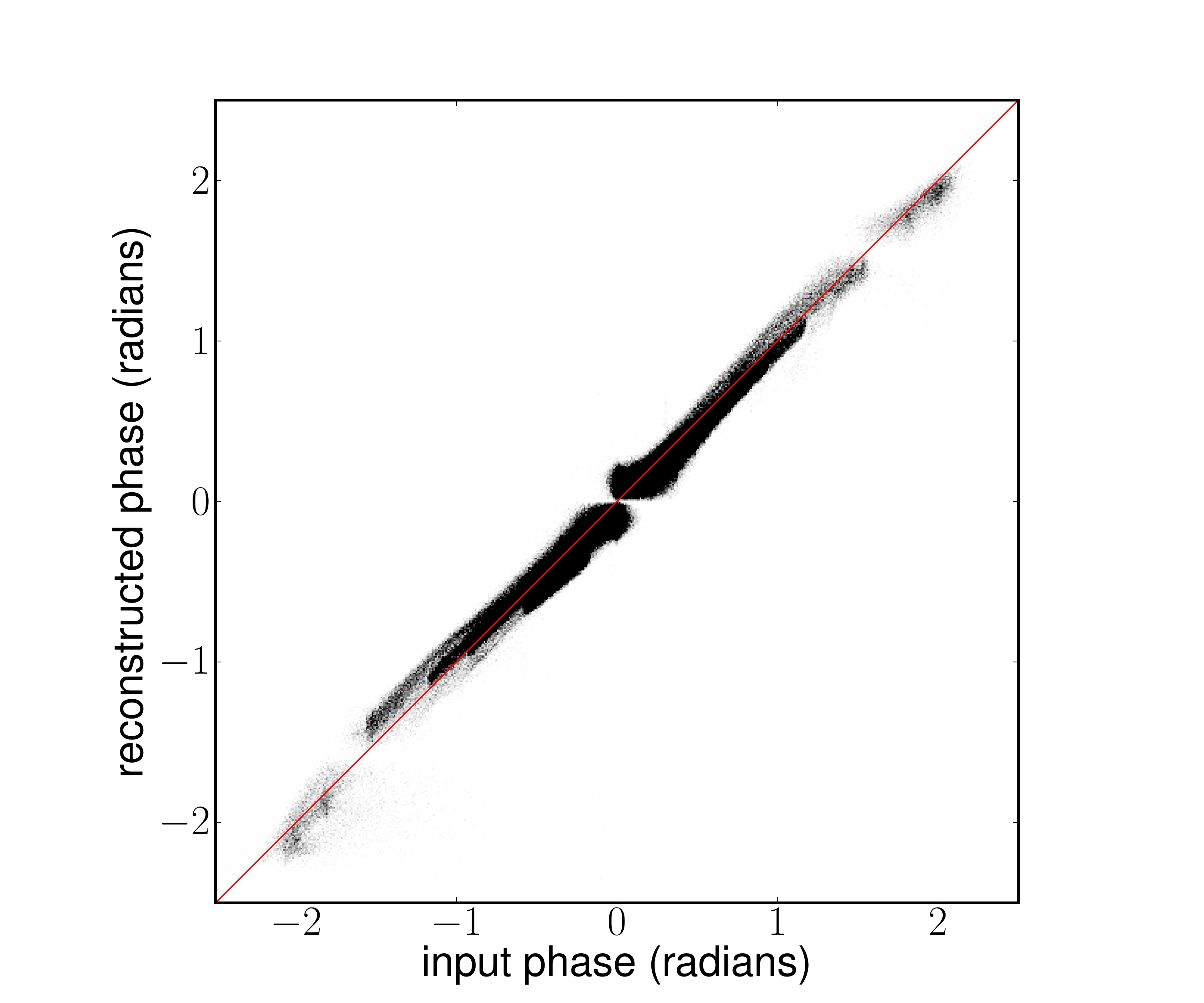}}
\caption{Plot of unwrapped reconstructed phase versus the input phase. The red line shows a perfect reconstruction.}
\label{fig:zigzag}
\end{figure}

Figure \ref{fig:zigzag} shows the reconstructed phase versus the input phase for the dataset taken at 750 nm.
This image shows that the reconstructed phase matches the input phase to within $0.1$ radians.
At an input phase of $\phi=0$ and $\phi=\pm \pi/2$ radians the reconstructed phase shows higher dispersion. 
This can be seen by the slightly flaring datapoints at those locations and is expected from simulations made with an equal amount of noise for the individual images. 
The measurement noise on the unbinned raw images at 750 nm propagates to an error on the phase per pixel of $0.07$ radians.

Adding Poisson noise systematically offsets the reconstructed retardance values.
To investigate how much this influences our reconstruction, we generate images with $\delta = \pi$ radians, rescale  the intensity images with the measured flux of the images in the lab, and inject a similar amount of Gaussian noise.
We obtain a systematically different average value across the pupil instead of the input half-wave retardance. 
This systematic offset becomes more significant as the noise is increased, and is an indicator of the noise limit on the retardance, as plotted in Fig.~\ref{fig:measured_retardance}.

Our images highly oversample (approximately by a factor of 4)  the spatial scales in the vAPP. 
To mitigate the Poisson noise, we bin the data before performing the reconstruction without loss of spatial information. 
If there is no significant change of reconstructed retardance as a function of binning, then we conclude that the results are not limited by Poisson noise. 
Using $4\times4$ binned simulated data we derive a measurement noise per pixel of $0.02$ radians on the phase and $0.08$ radians on the retardance using the measurement noise at 750 nm. 
At shorter wavelengths the noise level per pixel improves.

\section{PSF modeling theory}
\label{sec:appb}
To model the coronagraphic PSFs, we need to compute the Jones vector for both fully polarized beams, as it retaines the complete phase information that determines the destructive interference inside the dark hole.
The vAPP itself is modeled with a Jones matrix of an ideal retarder with a retardance $\delta_\mathrm{HWP}$ and fast axis $\theta_\mathrm{HWP}$. 
Ideally the retardance of the half-wave plate $\delta_\mathrm{HWP}=\pi$ radians and the fast axis $\theta_\mathrm{HWP}=\phi_\mathrm{APP}/2$, where $\phi_\mathrm{APP}$ is the true phase pattern in the pupil.
The Jones matrix of the HWP with fast axis $\theta_\mathrm{HWP}=0$ radians therefore is 
\begin{equation}
\mathbf{J_\mathrm{HWP,0}}=\left( \begin{matrix} 
\exp(-i\, \delta_\mathrm{HWP}/2) & 0 \\ 
 0 & \exp(i\, \delta_\mathrm{HWP}/2)
\end{matrix} \right).
\end{equation}
This matrix is rotated to the correct fast axis using
\begin{equation}
\mathbf{J_\mathrm{HWP,\theta}}=\mathbf{T}(-\theta)\, \mathbf{J_\mathrm{HWP,0}} \, \mathbf{T}(\theta)
\end{equation} for every pixel of the half-wave plate, where
\begin{equation}
\mathbf{T}(\theta)=\left( \begin{matrix} 
\cos{\theta} & \sin{\theta} \\
-\sin{\theta}& \cos{\theta}
\end{matrix} \right)\end{equation} is the rotation matrix with angle of rotation $\theta$.

The second optical component in our setup is a quarter-wave plate that converts the two circular polarizations into orthogonal linear polarization states. 
The final contrast also depends on the quality and alignment of this optic. 
\noindent The QWP can be described in a similar fashion as the HWP:
\begin{equation}
\mathbf{J_\mathrm{QWP,0}}=\left( \begin{matrix} 
\exp(-i\, \delta_\mathrm{QWP}/2) & 0 \\ 
 0 & \exp(i\, \delta_\mathrm{QWP}/2) 
\end{matrix} \right).
\end{equation}
The value of $\delta_\mathrm{QWP}$ should be $\pi/2$ in the ideal case. 
This matrix is rotated by a value $\theta_\mathrm{QWP}$ which ideally is $\pi/4$ to fully convert circular into linear polarization at 0,90$^\circ$ yielding
\begin{equation}
\mathbf{J_\mathrm{QWP,\theta}}=\mathbf{T}(-\theta)\, \mathbf{J_\mathrm{QWP,0}} \, \mathbf{T}(\theta).
\end{equation}

The final element in our Jones model is the Wollaston prism which can be described by a linear polarizer with an efficiency $\epsilon_\mathrm{pol}$.
The corresponding Jones matrix for one beam of the Wollaston is
\begin{equation}
\mathbf{J_\mathrm{pol,\perp}}=\left( \begin{matrix} 
\epsilon_\mathrm{pol} & 0 \\
0& 1-\epsilon_\mathrm{pol}
\end{matrix} \right).\\\end{equation}
The other beam can be described in a similar way with 
\begin{equation}
\mathbf{J_\mathrm{pol,\parallel}}=\left( \begin{matrix} 
1-\epsilon_\mathrm{pol} & 0 \\
0& \epsilon_\mathrm{pol}
\end{matrix} \right).\\\end{equation}
A rotation offset of the Wollaston also introduces leakage terms, and therefore these Jones matrices are also rotated with a rotation matrix.
The angle of rotation $\theta_\mathrm{pol}=0$ radians if placed perfectly but in practice can vary by a few degrees.

After defining these three optical components we study the effect of the optical setup on circular polarization input states as defined in Eq. \eqref{eq:circpol}.
The different components are combined in matrix form using
\begin{equation}
\mathbf{E_\mathrm{out}}= \mathbf{J_\mathrm{pol}} \, \mathbf{J_\mathrm{QWP,\frac{\pi}{4}}} \, \mathbf{J_\mathrm{HWP,\theta_\mathrm{APP}}} \, \mathbf{E_\mathrm{input}}.
\label{eq:PSF}
\end{equation}

\begin{figure}[h]
\centerline{\includegraphics[scale=0.25]{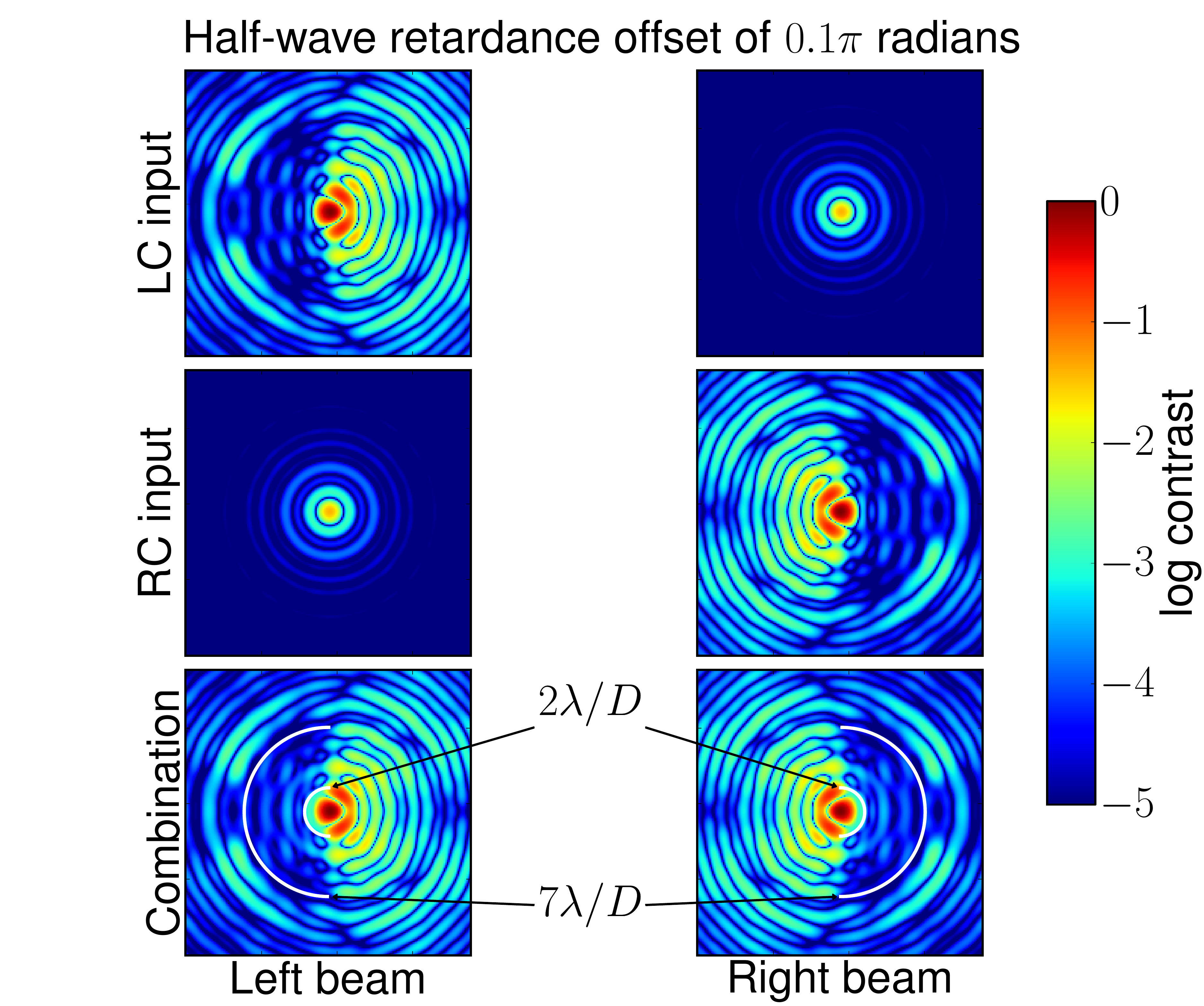}}
\caption{Effect of retardance offset of the HWP on the PSFs. Any offsets of the half-wave retardance of the vAPP fill the dark hole with an attenuated copy of the original PSF.}
\label{fig:hwpoffset}
\end{figure}

With no vAPP in the laboratory setup, Left circularly (LHC) polarized light will form an Airy PSF in only one of the two Wollaston beams, which we label as the Left beam and the Right beam. In our definition, the LHC polarized light will form an Airy pattern in the Right beam.
Inserting an ideal vAPP will then remove the Airy PSF from the Right beam and form the vAPP PSF in the Left beam.
Introducing right circularly (RHC) polarized light will produce the complementary vAPP PSF in the Right beam.
Recall that the vAPP is a patterned half-wave plate. 
If the optic is not exactly half-wave, then part of the flux from the vAPP appears as a leakage of the original Airy PSF in the other Wollaston beam. 
This effect is shown for a retardance offset of $0.1\pi$ radians in Fig. \ref{fig:hwpoffset}. 
The strength of the leakage increases with the offset from half-wave.

\begin{figure}[h]
\centerline{\includegraphics[scale=0.25]{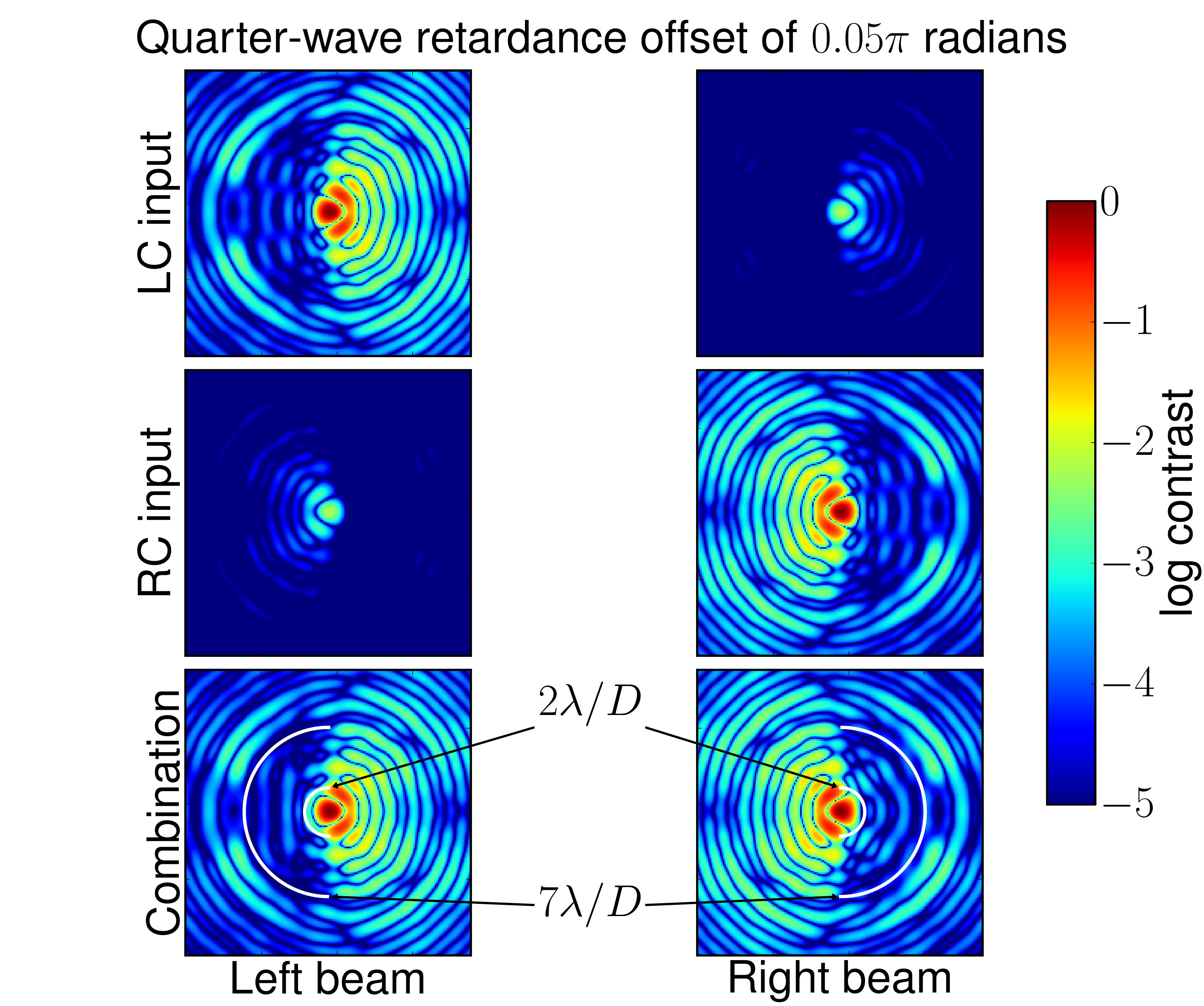}}
\caption{Effect of retardance offset of the QWP on the PSFs. Retardance offsets of the QWP fill the dark hole with the opposite handed PSF. Note that the leakage is normalized to the non-leaking term contribution to the PSF.}
\label{fig:qwpoffset}
\end{figure}

Another leakage term appears when the QWP departs from quarter-wave retardance.
 If the vAPP is illuminated with LHC polarized light, the leakage term appears as an attenuated APP PSF in the Right beam. 
 For a retardance offset of $0.05\pi$ radians this is shown in Fig. \ref{fig:qwpoffset}.

The other error contributions in the system include offsets from the desired phase pattern, misalignments in the QWP and Wollaston, and additional aberrations of the wavefront, which modify the phase pattern and the resultant PSFs, in a different way.
Also transmission variations in the vAPP and in the other optics change the PSF. 
We model all these effects with the Jones formalism and Fourier transform the resultant complex phase to model the PSF.






\begin{thebibliography}{10}
\newcommand{\enquote}[1]{``#1''}

\bibitem{Wolszczan:92}
A.~{Wolszczan} and D.~A. {Frail}, \enquote{{A planetary system around the
  millisecond pulsar PSR1257 + 12},} \nat \textbf{355}, 145--147 (1992).

\bibitem{Mayor:95}
M.~{Mayor} and D.~{Queloz}, \enquote{{A Jupiter-mass companion to a solar-type
  star},} \nat \textbf{378}, 355--359 (1995).

\bibitem{Charbonneau:00}
D.~{Charbonneau}, T.~M. {Brown}, D.~W. {Latham}, and M.~{Mayor},
  \enquote{{Detection of planetary transits across a Sun-like star},} \apjl
  \textbf{529}, L45--L48 (2000).

\bibitem{Borucki:09}
W.~{Borucki}, D.~{Koch}, N.~{Batalha}, D.~{Caldwell},
  J.~{Christensen-Dalsgaard}, W.~D. {Cochran}, E.~{Dunham}, T.~N. {Gautier},
  J.~{Geary}, R.~{Gilliland}, J.~{Jenkins}, H.~{Kjeldsen}, J.~J. {Lissauer},
  and J.~{Rowe}, \enquote{{KEPLER: Search for earth-size planets in the
  habitable zone},} in \enquote{IAU Symposium,} , vol. 253 of \emph{IAU
  Symposium}, F.~{Pont}, D.~{Sasselov}, and M.~J. {Holman}, eds. (2009), vol.
  253 of \emph{IAU Symposium}, pp. 289--299.

\bibitem{Batalha:13}
N.~M. {Batalha}, J.~F. {Rowe}, S.~T. {Bryson}, T.~{Barclay}, C.~J. {Burke},
  D.~A. {Caldwell}, J.~L. {Christiansen}, F.~{Mullally}, S.~E. {Thompson},
  T.~M. {Brown}, A.~K. {Dupree}, D.~C. {Fabrycky}, E.~B. {Ford}, J.~J.
  {Fortney}, R.~L. {Gilliland}, H.~{Isaacson}, D.~W. {Latham}, G.~W. {Marcy},
  S.~N. {Quinn}, D.~{Ragozzine}, A.~{Shporer}, W.~J. {Borucki}, D.~R. {Ciardi},
  T.~N. {Gautier}, III, M.~R. {Haas}, J.~M. {Jenkins}, D.~G. {Koch}, J.~J.
  {Lissauer}, W.~{Rapin}, G.~S. {Basri}, A.~P. {Boss}, L.~A. {Buchhave}, J.~A.
  {Carter}, D.~{Charbonneau}, J.~{Christensen-Dalsgaard}, B.~D. {Clarke}, W.~D.
  {Cochran}, B.-O. {Demory}, J.-M. {Desert}, E.~{Devore}, L.~R. {Doyle}, G.~A.
  {Esquerdo}, M.~{Everett}, F.~{Fressin}, J.~C. {Geary}, F.~R. {Girouard},
  A.~{Gould}, J.~R. {Hall}, M.~J. {Holman}, A.~W. {Howard}, S.~B. {Howell},
  K.~A. {Ibrahim}, K.~{Kinemuchi}, H.~{Kjeldsen}, T.~C. {Klaus}, J.~{Li}, P.~W.
  {Lucas}, S.~{Meibom}, R.~L. {Morris}, A.~{Pr{\v s}a}, E.~{Quintana}, D.~T.
  {Sanderfer}, D.~{Sasselov}, S.~E. {Seader}, J.~C. {Smith}, J.~H. {Steffen},
  M.~{Still}, M.~C. {Stumpe}, J.~C. {Tarter}, P.~{Tenenbaum}, G.~{Torres},
  J.~D. {Twicken}, K.~{Uddin}, J.~{Van Cleve}, L.~{Walkowicz}, and W.~F.
  {Welsh}, \enquote{{Planetary candidates observed by Kepler. III. Analysis of
  the first 16 months of data},} \apjs \textbf{204}, 24 (2013).

\bibitem{Marois:08}
C.~Marois, B.~Macintosh, T.~Barman, B.~Zuckerman, I.~Song, J.~Patience,
  D.~Lafrenière, and R.~Doyon, \enquote{Direct imaging of multiple planets
  orbiting the star HR 8799,} Science \textbf{322}, 1348--1352 (2008).

\bibitem{Marois:10}
C.~{Marois}, B.~{Zuckerman}, Q.~M. {Konopacky}, B.~{Macintosh}, and
  T.~{Barman}, \enquote{{Images of a fourth planet orbiting HR 8799},} \nat
  \textbf{468}, 1080--1083 (2010).

\bibitem{Oppenheimer:13}
B.R.~{Oppenheimer}, C.~{Baranec}, C.~{Beichman}, D.~{Brenner}, R.~{Burruss}, E.~{Cady}, J.R.~{Crepp}, R.~{Dekany}, R.~{Fergus}, D.~{Hale}, L.~{Hillenbrand}, S.~{Hinkley}, D.W.~{Hogg}, D.~{King}, E.R.~{Ligon}, T.~{Lockhart}, R.~{Nilsson}, I.R.~{Parry}, L.~{Pueyo}, E.~{Rice}, J.E.~{Roberts}, L.C.~{Roberts, Jr.}, M.~{Shao}, A.~{Sivaramakrishnan}, R.~{Soummer}, T.~{Truong}, G.~{Vasisht}, A.~{Veicht}, F.~{Vescelus}, J.K.~{Wallace}, C.~{Zhai} and N.~{Zimmerman}, \enquote{{Reconnaissance of the HR 8799 Exosolar System. I. Near-infrared Spectroscopy},} \apj \textbf{768}, 24 (2013).

\bibitem{Konopacky:13}
Q.~M. {Konopacky}, T.~S. {Barman}, B.~{Macintosh}, and C.~{Marois},
  \enquote{{Carbon and Oxygen in the spectrum of HR 8799c},} in
  \enquote{American Astronomical Society Meeting Abstracts \#221 p. 126.03.,} (2013). 

\bibitem{Marois:06}
C.~{Marois}, D.~{Lafreni{\`e}re}, R.~{Doyon}, B.~{Macintosh}, and D.~{Nadeau},
  \enquote{{Angular Differential Imaging: A powerful high-contrast imaging
  technique},} \apj \textbf{641}, 556--564 (2006).

\bibitem{Lafreniere:07}
D.~{Lafreni{\`e}re}, C.~{Marois}, R.~{Doyon}, D.~{Nadeau}, and
  {\'E}.~{Artigau}, \enquote{{A new algorithm for Point-Spread Function
  subtraction in high-contrast imaging: A demonstration with Angular
  Differential Imaging},} \apj \textbf{660}, 770--780 (2007).

\bibitem{Amara:12}
A.~{Amara} and S.~P. {Quanz}, \enquote{{PYNPOINT: an image processing package
  for finding exoplanets},} \mnras \textbf{427}, 948--955 (2012).

\bibitem{Meshkat:14}
T.~{Meshkat}, M.~A. {Kenworthy}, S.~P. {Quanz}, and A.~{Amara},
  \enquote{Optimized principal component analysis on coronagraphic images of
  the Fomalhaut System,} \apj \textbf{780}, 17 (2014).
  
  \bibitem{IRDIS}
 A.~ {Vigan}, C.~{Moutou}, M.~{Langlois}, D.~{Mouillet}, K.~{Dohlen}, A.~{Boccaletti}, M.~{Carbillet}, I.~{Smith}, A.~{Ferrari}, L.~{Mugnier}, and C.~{Thalmann}, \enquote{Comparison of methods for detection and characterization of exoplanets with SPHERE/IRDIS}, Proc.~SPIE 7735, 77352X (2011).
  
  \bibitem{SparksFord}
  W.~B.~Sparks, and H.~C.~Ford, \enquote{Imaging Spectroscopy for Extrasolar Planet Detection}, \apj \textbf{578}, pp.~543-564 (2002).

\bibitem{Avenhaus:14}
H.~Avenhaus, S.~P. Quanz, M.~R. Meyer, S.~D. Brittain, J.~S. Carr, and J.~R.
  Najita, \enquote{HD100546 multi-epoch scattered light observations,} \apj
  \textbf{790}, 56 (2014).

\bibitem{Bloemhof:03}
E.E.~{Bloemhof}, \enquote{{Suppression of speckle noise by speckle pinning in Adaptive Optics},} \apjl \textbf{582}, L59 (2003).

\bibitem{Lyot:39}
B.~{Lyot}, \enquote{{The study of the solar corona and prominences without
  eclipses (George Darwin Lecture, 1939)},} \mnras \textbf{99}, 580 (1939).

\bibitem{Guyon:05}
O.~{Guyon}, \enquote{{Limits of Adaptive Optics for high-contrast imaging},}
  \apj \textbf{629}, 592--614 (2005).

\bibitem{Mawet:12}
D.~{Mawet}, L.~{Pueyo}, P.~{Lawson}, L.~{Mugnier}, W.~{Traub}, A.~{Boccaletti},
  J.~T. {Trauger}, S.~{Gladysz}, E.~{Serabyn}, J.~{Milli}, R.~{Belikov},
  M.~{Kasper}, P.~{Baudoz}, B.~{Macintosh}, C.~{Marois}, B.~{Oppenheimer},
  H.~{Barrett}, J.-L. {Beuzit}, N.~{Devaney}, J.~{Girard}, O.~{Guyon},
  J.~{Krist}, B.~{Mennesson}, D.~{Mouillet}, N.~{Murakami}, L.~{Poyneer},
  D.~{Savransky}, C.~{V{\'e}rinaud}, and J.~K. {Wallace}, \enquote{{Review of
  small-angle coronagraphic techniques in the wake of ground-based
  second-generation adaptive optics systems},}  Proc. SPIE 8442, 844204 (2012).

\bibitem{Quanz:10}
S.P.~Quanz, M.R.~Meyer, M.A.~Kenworthy, J.H.V~Girard, M.~Kasper, A.-M.~Lagrange, D.~Apai, A.~Boccaletti, M.~Bonnefoy, G.~Chauvin, P.M.~Hinz and R.~Lenzen,\enquote{{First results from Very Large Telescope NACO Apodizing Phase Plate: 4 $\mu$m images of the exoplanet β Pictoris b},} \apjl \textbf{722}, L49 (2010).

\bibitem{Mawet:13}
D.~{Mawet}, O.~{Absil}, C.~{Delacroix}, J.H.~{Girard}, J.~{Milli}, J.~{O’Neal}, P.~{Baudoz}, A.~{Boccaletti}, P.~{Bourget}, V.~{Christiaens}, P.~{Forsberg}, F.~{Gonte}, S.~{Habraken}, C.~{Hanot}, M.~{Karlsson}, M.~{Kasper}, J.-L.~{Lizon}, K.~{Muzic}, R.~{Olivier}, E.~{Peña}, N.~{Slusarenko}, L.E.~{Tacconi-Garman} and J.~{Surdej}, \enquote{{L’-band AGPM vector vortex coronagraph’s first light on VLT/NACO},} \aa \textbf{552}, L13 (2013).


\bibitem{Kasdin:04}
N.J.~{Kasdin}, R.J.~{Vanderbei}, M.G.~{Littman}, M.~{Carr} and D.N.~{Spergel}, \enquote{The shaped pupil coronagraph for planet finding coronagraphy: optimization, sensitivity, and laboratory testing,} Proc. SPIE 5487, 1312 (2004).

\bibitem{Carlotti:13}
A.~{Carlotti}, \enquote{{Apodized phase mask coronagraphs for arbitrary
  apertures},} \aap \textbf{551}, A10 (2013).

\bibitem{Codona:04}
J.~L. {Codona} and R.~{Angel}, \enquote{{Imaging extrasolar planets by stellar
  halo suppression in separately corrected color bands},} \apjl \textbf{604},
  L117--L120 (2004).

\bibitem{Yang:03}
W.~{Yang} and A.B.~{Kostinski},\enquote{{Phase-modulated pupil for achromatic imaging of faint companions},} Phys. Lett. \textbf{320},  5 (2003).

\bibitem{Yang:04}
W.~{Yang} and A.B.~{Kostinski},\enquote{{One-sided achromatic phase apodization for imaging of extrasolar planets},} \apj \textbf{605}, 892 (2004). 

\bibitem{Guyon:05b}
O.~{Guyon}, E.A.~{Pluzhnik}, R.~{Galicher}, F.~{Martinache}, S.T.~{Ridgway} and R.A.~{Woodruff}, \enquote{{Exoplanet imaging with a phase-induced amplitude apodization coronagraph. I. Principle},} \apj \textbf{622}, 744 (2005).

\bibitem{Kostinski:05}
A.B.~{Kostinski} and W.~{Yang},\enquote{{Pupil phase apodization for imaging of faint companions in prescribed regions},} J. Mod. Opt 52, 2467--2474  (2005).

\bibitem{Kenworthy:07}
M.~A. {Kenworthy}, J.~L. {Codona}, P.~M. {Hinz}, J.~R.~P. {Angel}, A.~{Heinze},
  and S.~{Sivanandam}, \enquote{{First On-Sky High-Contrast Imaging with an
  Apodizing Phase Plate},} \apj \textbf{660}, 762--769 (2007).

\bibitem{Codona:06}
J.~L. {Codona}, M.~A. {Kenworthy}, P.~M. {Hinz}, J.~R.~P. {Angel}, and N.~J.
  {Woolf}, \enquote{{A high-contrast coronagraph for the MMT using phase
  apodization: design and observations at 5 microns and 2 {$\lambda$}/D
  radius},} Proc. SPIE 6269, 62691N (2006).

\bibitem{Yaroshchuk:12}
O.~Yaroshchuk and Y.~Reznikov, \enquote{Photoalignment of liquid crystals:
  basics and current trends,} J. Mater. Chem. \textbf{22}, 286--300 (2012).

\bibitem{Miskiewicz:14}
M.~N. Miskiewicz and M.~J. Escuti, \enquote{Direct-writing of complex liquid
  crystal patterns,} Opt. Express \textbf{22}, 12691--12706 (2014).

\bibitem{Komanduri:13}
R.~K. Komanduri, K.~F. Lawler, and M.~J. Escuti, \enquote{Multi-twist
  retarders: broadband retardation control using self-aligning reactive liquid
  crystal layers,} Opt. Express \textbf{21}, 404--420 (2013).

\bibitem{Pancharatnam:56}
S.~Pancharatnam, \enquote{Generalized theory of interference, and its
  applications. part i. coherent pencils,} in \enquote{Proceedings of the
  Indian Academy of Sciences, Section A,}, (Indian Academy of
  Sciences), vol.~44, pp. 247--262, (1956).

\bibitem{Berry:87}
M.~Berry, \enquote{The adiabatic phase and Pancharatnam's phase for polarized
  light,} J. Mod. Opt. \textbf{34}, 1401--1407 (1987).

\bibitem{Mawet:06}
D.~{Mawet}, P.~{Riaud}, J.~{Baudrand}, P.~{Baudoz}, A.~{Boccaletti},
  O.~{Dupuis}, and D.~{Rouan}, \enquote{{The four-quadrant phase-mask
  coronagraph: white light laboratory results with an achromatic device},} \aap
  \textbf{448}, 801--808 (2006).

\bibitem{Murakami:10}
N.~Murakami, J.~Nishikawa, K.~Yokochi, M.~Tamura, N.~Baba, and L.~Abe,
  \enquote{Achromatic eight-octant phase-mask coronagraph using photonic
  crystal,} \apj \textbf{714}, 772 (2010).
	
\bibitem{Mawet:09}
D.~{Mawet}, E.~{Serabyn}, K.~{Liewer}, C.~{Hanot}, S.~{McEldowney}, D.~{Shemo},
  and N.~{O'Brien}, \enquote{{Optical Vectorial Vortex Coronagraphs using
  Liquid Crystal Polymers: theory, manufacturing and laboratory
  demonstration},} Opt. Express \textbf{17}, 1902--1918 (2009).

\bibitem{Snik:12}
F.~{Snik}, G.~{Otten}, M.~{Kenworthy}, M.~{Miskiewicz}, M.~{Escuti},
  C.~{Packham}, and J.~{Codona}, \enquote{{The vector-APP: a broadband
  apodizing phase plate that yields complementary PSFs},} Proc. SPIE 8450, 84500M (2012).
  
  \bibitem{Snik:14spie}
F.~{Snik}, G.~{Otten}, M.~{Kenworthy}, D.~{Mawet}, and M.~{Escuti},
  \enquote{Combining vector-phase coronagraphy with dual-beam polarimetry,} Proc. SPIE 9147, 91477U (2014).

\bibitem{Carlotti:13b}
A.~Carlotti, N.~J. Kasdin, R.~J. Vanderbei, and A.~J.~E. Riggs, \enquote{Hybrid
  coronagraphic design: optimization of complex apodizers,} Proc. SPIE 8864, 88641Q (2013).

\bibitem{Otten:14spie}
G.~{Otten}, F.~{Snik}, M.~{Kenworthy}, J.~{Codona}, M.~{Escuti}, and
  M.~{Miskiewicz}, \enquote{Vector apodizing phase plate coronagraph:
  prototyping, characterization and outlook,} Proc. SPIE 9151, 91511R (2014).


\bibitem{Packham:10}
C.~Packham, M.~Escuti, J.~Ginn, C.~Oh, I.~Quijano, and G.~Boreman,
  \enquote{Polarization gratings: A novel polarimetric component for
  astronomical instruments,} Publications of the Astronomical Society of the
  Pacific \textbf{122}, pp. 1471--1482 (2010).

\bibitem{Roelfsema:10}
R.~Roelfsema, H.~M. Schmid, J.~Pragt, D.~Gisler, R.~Waters, A.~Bazzon,
  A.~Baruffolo, J.-L. Beuzit, A.~Boccaletti, J.~Charton, C.~Cumani, K.~Dohlen, M.~Downing, E.~Elswijk, M.~Feldt, C.~Groothuis, M.~de Haan, H.~Hanenburg, N.~Hubin, F.~Joos, M.~Kasper, C.~Keller, J.~Kragt, J.-L.~Lizon, D.~Mouillet, A.~Pavlov, F.~Rigal, S.~Rochat, B.~Salasnich, P.~Steiner, C.~Thalmann, L.~Venema, F.~Wildi,
  \enquote{The ZIMPOL high-contrast imaging polarimeter for sphere: design,
  manufacturing, and testing,} Proc. SPIE 7735, 77354B (2010).

\bibitem{Macintosh:14}
B.~Macintosh, J.~R. Graham, P.~Ingraham, Q.~Konopacky, C.~Marois, M.~Perrin,
  L.~Poyneer, B.~Bauman, T.~Barman, A.~S. Burrows, A.~Cardwell, J.~Chilcote,
  R.~J. De~Rosa, D.~Dillon, R.~Doyon, J.~Dunn, D.~Erikson, M.~P. Fitzgerald,
  D.~Gavel, S.~Goodsell, M.~Hartung, P.~Hibon, P.~Kalas, J.~Larkin, J.~Maire,
  F.~Marchis, M.~S. Marley, J.~McBride, M.~Millar-Blanchaer, K.~Morzinski,
  A.~Norton, B.~R. Oppenheimer, D.~Palmer, J.~Patience, L.~Pueyo, F.~Rantakyro,
  N.~Sadakuni, L.~Saddlemyer, D.~Savransky, A.~Serio, R.~Soummer,
  A.~Sivaramakrishnan, I.~Song, S.~Thomas, J.~K. Wallace, S.~Wiktorowicz, and
  S.~Wolff, \enquote{First light of the Gemini Planet Imager,} Proceedings of
  the National Academy of Sciences (2014).

\bibitem{Rodenhuis:12}
M.~{Rodenhuis}, H.~{Canovas}, S.~V. {Jeffers}, M.~{de Juan Ovelar}, M.~{Min},
  L.~{Homs}, and C.~U. {Keller}, \enquote{{The extreme polarimeter: design,
  performance, first results and upgrades},} Proc. SPIE 8446, 84469I (2012).

\bibitem{Packham:12}
C.~Packham, T.~J. Jones, C.~Warner, M.~Krejny, D.~Shenoy, T.~Vonderharr,
  E.~Lopez-Rodriguez, and K.~DeWahl, \enquote{Commissioning results of MMT-POL:
  the 1-5um imaging polarimeter leveraged from the AO secondary of the 6.5m
  MMT,} Proc. SPIE 8446, 84463R (2012).

\bibitem{Skrutskie:10}
M.~F. {Skrutskie}, T.~{Jones}, P.~{Hinz}, P.~{Garnavich}, J.~{Wilson},
  M.~{Nelson}, E.~{Solheid}, O.~{Durney}, W.~{Hoffmann}, V.~{Vaitheeswaran},
  T.~{McMahon}, J.~{Leisenring}, and A.~{Wong}, \enquote{{The Large Binocular
  Telescope mid-infrared camera (LMIRcam): final design and status},} Proc. SPIE 7735, 77353H (2010).

\bibitem{Riaud:12}
P.~Riaud, D.~Mawet, and A.~Magette, \enquote{Instantaneous phase retrieval with the vector vortex coronagraph. Theoretical and optical implementation,} A\&A \textbf{545}, A151 (2012).
  

\end{thebibliography}
\end{document}